\documentclass[prd,showkeys,floatfix,twocolumn,amsmath,amssymb]{revtex4}
\usepackage{graphicx,color,dcolumn,booktabs,bm}
\usepackage{longtable,lscape}
\usepackage{indentfirst}
\usepackage{epsfig}
\usepackage{feynmf}
\usepackage{epstopdf}
\usepackage{slashed}
\usepackage{cases}
\usepackage{multirow}
\usepackage[colorlinks,citecolor=blue,anchorcolor=red,menucolor=red,linkcolor=red,filecolor=red,runcolor=red,urlcolor=blue]{hyperref}

\newcommand{\feynp}[1]{#1\kern-0.45em/}

\def\FF(s){\left[(\alpha+\beta)m_c^2-\alpha\beta s\right]}
\def\HH(s){\left[m_c^2-\alpha(1-\alpha) s\right]}
\def\KK(s){\left[\gamma m_c^2-\gamma(1-\gamma) s\right]}

\allowdisplaybreaks[3]

\begin{document}
\title{Exploring a low-mass H dibaryon in light of the LZ event: A QCD sum rule study}
%

\author{Wei-Han Tan}
\affiliation{School of Physics, Southeast University, Nanjing 211189, China}

\author{Hua-Xing Chen}
\email{hxchen@seu.edu.cn}
\affiliation{School of Physics, Southeast University, Nanjing 211189, China}

\begin{abstract}

The LUX-ZEPLIN (LZ) Collaboration recently reported
a candidate event consistent with a high-energy nuclear
recoil, motivating a possible interpretation by
Lee, Takahashi, and Tsai involving the production of
an invisible baryon-number-two scalar with a mass
near $1.85$~GeV.
Motivated by this hypothetical scenario, we investigate
the $J^P=0^+$ $uuddss$ hexaquark system using QCD sum rules.
We consider five local diquark--diquark--diquark
interpolating currents with different scalar and
axial-vector diquark combinations, together with a
baryon--baryon current constructed from two
$\Lambda$-type baryon currents.
The operator product expansions are calculated up to
dimension sixteen, and their numerical behaviors are
examined under different OPE truncations.
Among the currents considered, only the
$SU(3)_{\rm flavor}$-singlet current $J_{SSS}$,
consisting of three scalar color-$\bar{\mathbf{3}}_C$ diquarks,
yields a low-mass solution at $D=12$, with
$M_{SSS}=1.85^{+0.19}_{-0.20}$~GeV.
Interestingly, scalar color-$\bar{\mathbf{3}}_C$ diquarks
belong to the flavor $\bar{\mathbf{3}}_F$
representation, and their local color-singlet
construction has a unique nonvanishing flavor
structure, $[ud][ds][su]$, corresponding to $uuddss$.
Analogous all-scalar-diquark currents vanish for
other flavor configurations, such as $uuuddd$ and
$uuudds$. Together with our QCD sum-rule results,
this property highlights the special role of the
all-scalar-diquark configuration, $[ud][ds][su]$, as a
possible deeply bound structure.
However, the higher-dimensional OPE contributions
exhibit significant truncation dependence, and the
low-mass solution remains conditional on the adopted
OPE truncation rather than constituting a fully
controlled mass prediction.

\end{abstract}

\keywords{dibaryon, hexaquark, diquark correlations, QCD sum rules}
\maketitle
\pagenumbering{arabic}

\section{Introduction}
\label{sec:intro}

The possible existence of dibaryons provides an important arena for
studying the nonperturbative dynamics of the strong interaction beyond
the conventional meson and baryon sectors~\cite{pdg}. Among them, the $H$
dibaryon has attracted particular attention since it was first proposed
as a deeply bound six-quark state with the valence-quark content
$uuddss$~\cite{Jaffe:1976yi}. Its most commonly discussed quantum numbers
are
\begin{equation}
I=0,\qquad S=-2,\qquad J^P=0^+,
\end{equation}
so that it can couple to the $\Lambda\Lambda$, $N\Xi$, and
$\Sigma\Sigma$ channels. Depending on the underlying dynamics, such a
state may involve compact six-quark correlations, baryon--baryon
components, or a mixture of different configurations.

The $H$ dibaryon has been extensively investigated using various
theoretical approaches~\cite{Kochelev:1999pg,Suganuma:2016lmp,Clement:2020mab,Aoki:2023qih,Gal:2024nbr,Moore:2024mot,Moore:2025zyg,Miwa:2025adw,Zhang:2025qmg}. 
Lattice QCD calculations have found evidence
for attractive interactions in the flavor-singlet baryon--baryon
channel and, at unphysically large light-quark masses, for bound
$H$-dibaryon states below the $\Lambda\Lambda$
threshold~\cite{Inoue:2010es,NPLQCD:2010ocs,Green:2021qol}.
Near-physical-mass lattice studies have also examined the
coupled-channel interactions involving $\Lambda\Lambda$ and
$N\Xi$~\cite{HALQCD:2019wsz}.
More recently, Hu and Guo~\cite{Hu:2026hdb} investigated the
baryon--baryon interactions within an $SU(6)$-constrained effective
field theory and identified two bound-state poles below the
$\Lambda\Lambda$ threshold within their framework.
Constituent-quark-model calculations have also investigated
compact six-quark configurations in the $uuddss$ system,
but found no compact bound state within the adopted
interaction models~\cite{Park:2016cmg,Gordillo:2026usv}.
These investigations indicate that the possible existence
and internal structure of the $H$ dibaryon remain
important open questions.

A particularly interesting possibility is that the $uuddss$ system
forms a compact scalar hexaquark, or sexaquark, with nontrivial
color-spin-flavor correlations and hidden-color
components~\cite{Farrar:2023wvm}.
In particular, Gross et al.~\cite{Gross:2018ivp} and
Buccella~\cite{Buccella:2020mxi} investigated the
three-scalar-diquark configuration $[ud][ds][su]$,
with the latter obtaining a mass estimate near $1.88$~GeV.
A holographic QCD study also obtained a low-mass estimate
near $1.75$~GeV, depending on the assumed anomalous
dimensions~\cite{Evans:2023zde}.
Its production and possible experimental searches have also
been discussed~\cite{She:2025dqx,Doser:2023gls}.
Such a compact configuration may be particularly relevant
far below the $\Lambda\Lambda$ threshold, where a weakly
bound molecular description becomes inadequate.

Very recently, the LUX-ZEPLIN (LZ) Collaboration reported one event
with characteristics consistent with a nuclear recoil of
\begin{equation}
E_R=248\pm23_{\rm stat}\pm23_{\rm sys}~{\rm keV},
\end{equation}
in an extended nuclear-recoil search window~\cite{LZ:2026axp}.
The reported global significance is $2.6\sigma$,
which is insufficient to establish a new physical phenomenon.
Nevertheless, this event has motivated a variety of theoretical interpretations and related phenomenological studies~\cite{Yin:2026jnn,Fan:2026kxx,Wu:2026nhi,Freese:2026sga,Su:2026rwz,Yamashita:2026ump,DiMauro:2026ldr,Rodd:2026tyn,McCabe:2026crm,Du:2026guj,Unwin:2026rdp,Bose:2026ndd,Bisal:2026khf,Langhoff:2026ujr,Yuan:2026djt,Fan:2026hzw,Nguyen:2026lui,Tang:2026lgu,Gemmell:2026yaw,Nagata:2026pbj,An:2026pkc,Aghaie:2026vsu,Uttayarat:2026isp,Lian:2026hpm,Lian:2026bze,Das:2026buc,Lou:2026idn,Pallis:2026jdc,Borah:2026zwf,Paul:2026okh,Sahoo:2026dnw,Bose:2026szs,De:2026win,Chanda:2026wdl,Anwar:2026ffj,Lee:2026jxl,Lee:2026tro,CentellesChulia:2026ats,Das:2026uyy,Mahapatra:2026glu,Palmisano:2026kuj}.
In particular, Lee, Takahashi, and Tsai~\cite{Lee:2026zbr}
proposed that two bound neutrons inside a nucleus may convert
into an invisible scalar particle carrying baryon number two,
\begin{equation}
(A,Z)\rightarrow(A-2,Z)^{(*)}+\phi .
\end{equation}
For the transition
\begin{equation}
{}^{132}{\rm Xe}
\rightarrow
{}^{130}{\rm Xe}
+\phi ,
\end{equation}
assuming that the observed recoil arises from
a transition to the daughter ground state leads to
an invisible scalar mass near
\begin{equation}
m_\phi\simeq1.847~{\rm GeV}.
\label{eq:LZ_mass_intro}
\end{equation}
An intriguing possibility is to identify this
invisible scalar with a composite $B=2$ state
having the quark content $uuddss$ and $J^P=0^+$.
Such an identification remains hypothetical and
would require a strangeness-changing transition
with $\Delta S=-2$ mediated by weak interactions.

This hypothetical mass scale is substantially lower than
that usually considered for a weakly bound $H$ dibaryon.
Using $2M_\Lambda\simeq2.23$~GeV as the natural
baryon--baryon threshold, a state around $1.85$~GeV
would lie approximately
\begin{equation}
2M_\Lambda-M_S\simeq0.38~{\rm GeV}
\end{equation}
below the $\Lambda\Lambda$ threshold.
Such a large binding energy is difficult to accommodate
within conventional weakly bound molecular descriptions.
It is therefore interesting to investigate whether QCD
can accommodate such a low-mass state through
specific multiquark correlations.

On the other hand, the possible existence of such a low-mass
$uuddss$ state also raises important questions concerning the
stability of ordinary nuclei. For a mass near $1.85$~GeV,
transitions of nucleon pairs into the $uuddss$ state can be
energetically allowed in various nuclei, depending on the
corresponding nuclear separation energies. In particular,
the possible decay of the deuteron provides an important
constraint. Previous studies have investigated these
nuclear-stability constraints and suggested that such
transitions may be strongly suppressed by the
strangeness-changing weak interactions and by the small
overlap between compact six-quark and ordinary two-baryon
configurations~\cite{Farrar:2003qy,Farrar:2023wta}. 
Nevertheless, whether these suppressions
can accommodate the LZ candidate event while satisfying
existing nuclear-stability constraints remains an open
question. A quantitative investigation of the corresponding
transition amplitudes is beyond the scope of the present study.

Motivated by the possibility of such a low-mass state,
in the present study we investigate the $J^P=0^+$
$uuddss$ system using QCD sum rules~\cite{Shifman:1978bx,Shifman:1978by}.
Previous QCD sum-rule studies of this system have
employed both six-quark and baryon--baryon
interpolating currents~\cite{Larin:1985yt,Kodama:1994np,Azizi:2019xla}.
Here we compare several local interpolating currents
with different internal structures.
We first examine diquark--diquark--diquark configurations
constructed from color-$\bar{\mathbf{3}}_C$
scalar and axial-vector diquarks.
An interesting feature is that scalar diquarks with
$J^P=0^+$ and color $\bar{\mathbf{3}}_C$ belong
to the flavor $\bar{\mathbf{3}}_F$ representation.
For the three light flavors $u$, $d$, and $s$,
the antisymmetric color-singlet construction from
three such local scalar diquarks has a unique
nonvanishing flavor structure,
\begin{equation}
[ud][ds][su],
\end{equation}
corresponding to $uuddss$~\cite{Gross:2018ivp,Buccella:2020mxi}.
In contrast, axial-vector diquarks with $J^P=1^+$
belong to the flavor $\mathbf{6}_F$ representation,
allowing a wider variety of six-quark flavor configurations.
We compare the all-scalar-diquark configuration with
currents containing axial-vector diquarks to examine
their different sum-rule behaviors.
We further consider a local $\Lambda\Lambda$-type
current as a baryon--baryon reference.

This paper is organized as follows. In Sec.~\ref{sec:current},
we analyze the color structures of the six-quark system
and construct the diquark-based and baryon--baryon
interpolating currents. The QCD sum-rule formalism
and the operator product expansion are presented in
Sec.~\ref{sec:sumrule}. In Sec.~\ref{sec:numerical},
we perform the numerical analyses and compare the
results obtained from different currents, together
with previous theoretical studies. Finally, we
summarize our findings and discuss their physical
implications and limitations in Sec.~\ref{sec:summary}.

\section{Interpolating Currents}
\label{sec:current}

\begin{figure*}[htbp]
\centering
\includegraphics[width=0.80\textwidth]{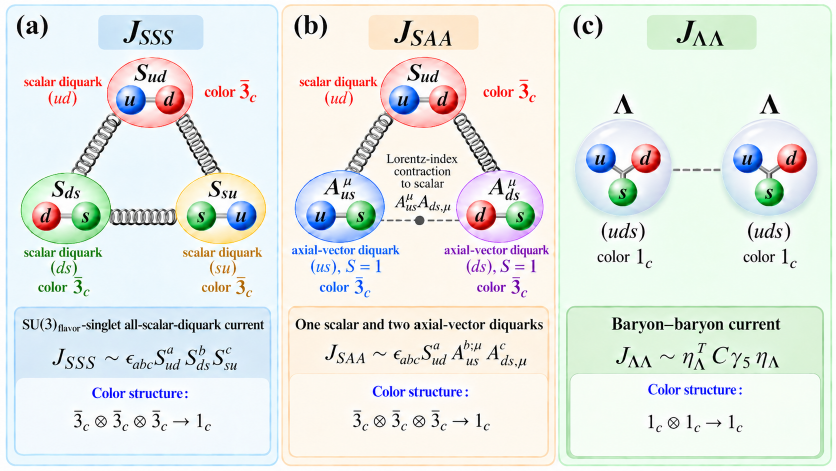}
\caption{
Schematic representations of three representative $uuddss$
interpolating currents considered in this work.
(a) The $SU(3)_{\rm flavor}$-singlet current $J_{SSS}$,
consisting of three scalar diquarks.
(b) The isoscalar current $J_{SAA}$, containing one scalar
and two axial-vector diquarks whose Lorentz indices are contracted
to form a scalar.
(c) The baryon--baryon current $J_{\Lambda\Lambda}$,
constructed from two local $\Lambda$-type three-quark currents.
The color structures are
$\bar{\mathbf{3}}_c\otimes\bar{\mathbf{3}}_c
\otimes\bar{\mathbf{3}}_c\to\mathbf{1}_c$
for (a) and (b), and
$\mathbf{1}_c\otimes\mathbf{1}_c\to\mathbf{1}_c$
for (c).
}
\label{fig:currents}
\end{figure*}

In this section, we construct local interpolating currents for the
$uuddss$ hexaquark system with $J^P=0^+$. Particular attention is
paid to the isoscalar and $SU(3)_{\rm flavor}$-singlet configurations
relevant to the $H$ dibaryon, while several auxiliary currents with
$I_3=0$ are also considered to investigate the dependence of the
QCD sum rules on different internal correlation structures.
The representative coupling schemes considered in the present work
are illustrated schematically in Fig.~\ref{fig:currents}. We consider
both diquark--diquark--diquark configurations and a baryon--baryon
configuration. Among the former, the $J_{SSS}$ current contains three
scalar diquarks, while $J_{SAA}$ contains one scalar and two
axial-vector diquarks. The $\Lambda\Lambda$ current is introduced as
a representative baryon--baryon configuration for comparison.

We first discuss the possible color structures of the six-quark
system, and then construct the corresponding diquark--diquark--diquark
and baryon--baryon interpolating currents.

\subsection{Color Structures}
\label{subsec:color}

We first consider the color structures of the
diquark--diquark--diquark configuration $[qq][qq][qq]$. Each quark
transforms as the fundamental representation $\mathbf{3}_c$ of the
color $SU(3)_{\rm color}$ group, and two quarks can be coupled either
to the antisymmetric color antitriplet or to the symmetric color
sextet,
\begin{equation}
\mathbf{3}_c\otimes\mathbf{3}_c
=
\bar{\mathbf{3}}_c\oplus\mathbf{6}_c .
\label{eq:color_qq}
\end{equation}
For a fixed partition of six quarks into three diquarks, an overall
color singlet can be obtained from the following five couplings:
\begin{eqnarray}
&&
\bar{\mathbf{3}}_c\otimes
\bar{\mathbf{3}}_c\otimes
\bar{\mathbf{3}}_c ,
\nonumber\\
&&
\bar{\mathbf{3}}_c\otimes
\bar{\mathbf{3}}_c\otimes
\mathbf{6}_c ,
\nonumber\\
&&
\bar{\mathbf{3}}_c\otimes
\mathbf{6}_c\otimes
\bar{\mathbf{3}}_c ,
\nonumber\\
&&
\mathbf{6}_c\otimes
\bar{\mathbf{3}}_c\otimes
\bar{\mathbf{3}}_c ,
\nonumber\\
&&
\mathbf{6}_c\otimes
\mathbf{6}_c\otimes
\mathbf{6}_c .
\label{eq:color_diquark}
\end{eqnarray}
These five couplings provide five independent color-singlet
structures for a fixed diquark--diquark--diquark partition.

Among these possibilities, the color-antitriplet diquark
$\bar{\mathbf{3}}_c$ provides a particularly natural starting point.
From the viewpoint of the one-gluon-exchange interaction, with
$T^A=\lambda^A/2$, the corresponding color factors are
\begin{equation}
\left\langle {\bf T}_1\cdot{\bf T}_2\right\rangle_{\bar{\mathbf{3}}_c}
=
-\frac{2}{3},
\qquad
\left\langle {\bf T}_1\cdot{\bf T}_2\right\rangle_{\mathbf{6}_c}
=
+\frac{1}{3}.
\label{eq:color_factor}
\end{equation}
The color-antitriplet channel is therefore attractive, whereas the
color-sextet channel is repulsive at the one-gluon-exchange level.
Motivated by this observation, in the present study we restrict the
diquark-based interpolating currents to those constructed from three
color-$\bar{\mathbf{3}}_c$ diquarks. Although color-$\mathbf{6}_c$
diquarks may also contribute in more general six-quark
configurations, such structures are not explicitly considered in the
present work.

Suppressing the spin and flavor structures, the color part of a
color-antitriplet diquark can be written, up to an overall
normalization, as
\begin{equation}
[qq]^a_{\bar{\mathbf{3}}_c}
=
\epsilon^{abc}\,q_b q_c ,
\label{eq:color_antitriplet}
\end{equation}
where $a,b,c$ denote color indices. Three such color-antitriplet
diquarks can then be coupled into an overall color singlet through
the color tensor
\begin{equation}
\epsilon_{abc}
\left[\epsilon^{a a_1 a_2}q_{a_1}q_{a_2}\right]
\left[\epsilon^{b b_1 b_2}q_{b_1}q_{b_2}\right]
\left[\epsilon^{c c_1 c_2}q_{c_1}q_{c_2}\right] .
\label{eq:color_three_diquark}
\end{equation}
Accordingly, all diquark-based currents considered in this work are
constructed from three color-$\bar{\mathbf{3}}_c$ diquarks.

We next consider the color structures of the baryon--baryon
configuration $[qqq][qqq]$. Three quarks can be coupled according to
\begin{equation}
\mathbf{3}_c\otimes\mathbf{3}_c\otimes\mathbf{3}_c
=
\mathbf{1}_c
\oplus
\mathbf{8}_c
\oplus
\mathbf{8}_c
\oplus
\mathbf{10}_c .
\label{eq:color_qqq}
\end{equation}
The two color-octet representations in Eq.~\eqref{eq:color_qqq}
correspond to two independent color structures arising from different
intermediate couplings of the three quarks. Consequently, for a fixed
partition of six quarks into two three-quark clusters, the
five-dimensional color-singlet subspace can be represented by one
$\mathbf{1}_c\otimes\mathbf{1}_c$ configuration and four independent
$\mathbf{8}_c\otimes\mathbf{8}_c$ configurations. The former
corresponds directly to two color-singlet baryons, whereas the latter
are usually referred to as hidden-color configurations.

Among these possibilities, the
$\mathbf{1}_c\otimes\mathbf{1}_c$ configuration provides the most
direct starting point for constructing baryon--baryon interpolating
currents, since each three-quark cluster can be directly associated
with a conventional color-singlet baryon. In the present study, we
restrict the baryon--baryon currents to this
$\mathbf{1}_c\otimes\mathbf{1}_c$ configuration. Although the
$\mathbf{8}_c\otimes\mathbf{8}_c$ hidden-color configurations may
play an important role in a compact six-quark system, they are not
explicitly included in the present analysis.

Suppressing the spin and flavor structures, the color part of a
color-singlet three-quark cluster can be written, up to an overall
normalization, as
\begin{equation}
[qqq]_{\mathbf{1}_c}
=
\epsilon^{abc}\,q_a q_b q_c ,
\label{eq:color_baryon}
\end{equation}
and two such color-singlet clusters form an overall color singlet
through
\begin{equation}
\left[\epsilon^{a_1a_2a_3}
q_{a_1}q_{a_2}q_{a_3}\right]
\left[\epsilon^{b_1b_2b_3}
q_{b_1}q_{b_2}q_{b_3}\right] .
\label{eq:color_two_baryons}
\end{equation}
Accordingly, all baryon--baryon currents considered in this work are
constructed from two color-singlet three-quark clusters.

The diquark--diquark--diquark and baryon--baryon constructions are
different coupling schemes of the same five-dimensional six-quark
color-singlet space, and their color structures can be transformed
into one another through $SU(3)_{\rm color}$ recoupling. For example,
the color structure of three color-antitriplet diquarks can be
explicitly rewritten as a linear combination of two products of
color-singlet three-quark clusters,
\begin{equation}
\epsilon_{abc}
\epsilon^{a a_1 a_2}
\epsilon^{b b_1 b_2}
\epsilon^{c c_1 c_2}
=
\epsilon^{a_1 b_1 b_2}
\epsilon^{a_2 c_1 c_2}
-
\epsilon^{a_2 b_1 b_2}
\epsilon^{a_1 c_1 c_2}.
\label{eq:color_recoupling}
\end{equation}
Conversely, baryon--baryon color structures can also be recoupled
into linear combinations of diquark--diquark--diquark structures.
When such color recoupling is combined with the corresponding Fierz
transformations in the Dirac and flavor spaces, local currents in
one coupling scheme can, within a complete local operator basis, be
expressed as linear combinations of currents in the other coupling
scheme. The two types of interpolating currents therefore probe
different coupling schemes and correlation structures of the same
underlying six-quark degrees of freedom, rather than representing
mutually independent physical components~\cite{Chen:2007xr,Chen:2016ymy}.

In practice, however, such complete transformations generally involve
a large number of color, Dirac, and flavor structures, making a full
coupled-current analysis considerably more complicated. In the
present study, we therefore restrict ourselves to several
representative individual currents with definite and physically
motivated color, Dirac, and flavor structures, and investigate the
corresponding QCD sum rules separately.

\subsection{Diquark--Diquark--Diquark Currents}
\label{subsec:diquark}

We first construct the $uuddss$ interpolating currents in the
diquark--diquark--diquark configuration. As discussed in
Sec.~\ref{subsec:color}, all diquark-based currents considered in the
present study are constructed from color-$\bar{\mathbf{3}}_c$
diquarks. For the local derivative-free diquark operators relevant
to our analysis, we consider the scalar and axial-vector diquark
structures,
\begin{eqnarray}
S_{q_1q_2}^{a}
&=&
\epsilon^{abc}
q_{1b}^{T} C\gamma_{5} q_{2c},
\label{eq:scalar_diquark}
\\
A_{q_1q_2}^{a;\mu}
&=&
\epsilon^{abc}
q_{1b}^{T} C\gamma^{\mu} q_{2c},
\label{eq:axial_diquark}
\end{eqnarray}
where $a,b,c$ are color indices and $C$ is the charge-conjugation
matrix.

The scalar diquark $S_{q_1q_2}^{a}$ carries $J^P=0^+$. In the
$S$-wave assignment adopted here,
\begin{equation}
S=J=0,
\qquad
L=0,
\qquad
P=(-1)^L=+,
\end{equation}
so that its orbital wave function is symmetric, whereas its spin wave
function is antisymmetric under the interchange of the two quarks.
Since its color wave function is also antisymmetric, the Pauli
principle requires the flavor wave function to be antisymmetric.
Accordingly, the scalar diquarks transform as the
$\bar{\mathbf{3}}_F$ representation of the flavor
$SU(3)_{\rm flavor}$ group,
\begin{equation}
\bar{\mathbf{3}}_F
=
\left\{
S_{ud}^{a},\,S_{ds}^{a},\,S_{su}^{a}
\right\}.
\label{eq:scalar_flavor}
\end{equation}
Equivalently,
\begin{equation}
S_{q_1q_2}^{a}
=
-S_{q_2q_1}^{a},
\qquad
S_{uu}^{a}=S_{dd}^{a}=S_{ss}^{a}=0.
\label{eq:scalar_flavor_symmetry}
\end{equation}

The axial-vector diquark $A_{q_1q_2}^{a;\mu}$ carries $J^P=1^+$.
In the same $S$-wave assignment,
\begin{equation}
S=J=1,
\qquad
L=0,
\qquad
P=(-1)^L=+,
\end{equation}
so that both its orbital and spin wave functions are symmetric under
the interchange of the two quarks. Since its color wave function is
antisymmetric, the Pauli principle requires the flavor wave function
to be symmetric. Accordingly, the axial-vector diquarks transform as
the $\mathbf{6}_F$ representation of the flavor
$SU(3)_{\rm flavor}$ group,
\begin{equation}
\mathbf{6}_F
=
\left\{
A_{uu}^{a;\mu},\,A_{dd}^{a;\mu},\,A_{ss}^{a;\mu},\,
A_{ud}^{a;\mu},\,A_{us}^{a;\mu},\,A_{ds}^{a;\mu}
\right\},
\label{eq:axial_flavor}
\end{equation}
with
\begin{equation}
A_{q_1q_2}^{a;\mu}
=
+A_{q_2q_1}^{a;\mu}.
\label{eq:axial_flavor_symmetry}
\end{equation}

A particularly constrained configuration is obtained when all three
diquarks are scalar. For the $uuddss$ flavor content, such a
configuration is fixed, up to permutations and an overall phase, as
\begin{equation}
(ud)_{0^+}(ds)_{0^+}(su)_{0^+}.
\end{equation}
The corresponding local interpolating current is
\begin{equation}
J_{SSS}
=
\epsilon_{abc}
S_{ud}^{a}
S_{ds}^{b}
S_{su}^{c},
\label{eq:JSSS}
\end{equation}
or, explicitly,
\begin{eqnarray}
J_{SSS}
&=&
\epsilon_{abc}
\epsilon^{aa_1a_2}
\epsilon^{bb_1b_2}
\epsilon^{cc_1c_2}
\nonumber\\
&&\times
\left(
u_{a_1}^{T}C\gamma_{5}d_{a_2}
\right)
\left(
d_{b_1}^{T}C\gamma_{5}s_{b_2}
\right)
\left(
s_{c_1}^{T}C\gamma_{5}u_{c_2}
\right).
\label{eq:JSSS_explicit}
\end{eqnarray}
This current transforms as an $SU(3)_{\rm flavor}$ singlet and hence
necessarily carries $I=0$. Within the $S$-wave assignment adopted
here, it has $J^P=0^+$. Its coupling to the
corresponding scalar state $|X_{SSS}(q)\rangle$ is defined by
\begin{equation}
\langle 0|J_{SSS}|X_{SSS}(q)\rangle
=
f_{SSS},
\label{eq:JSSS_coupling}
\end{equation}
where $f_{SSS}$ denotes the corresponding coupling constant. Thus,
the color, spin, and flavor structures of the all-scalar $uuddss$
current are highly constrained.

We further construct currents containing axial-vector diquarks in
order to investigate the dependence of the QCD sum-rule spectrum on
the internal spin correlations. For the $uuddss$ flavor content,
three local Lorentz-scalar currents containing one scalar and two
axial-vector diquarks are considered,
\begin{eqnarray}
J_{SAA}
&=&
\epsilon_{abc}
S_{ud}^{a}
A_{us}^{b;\mu}
A_{ds,\mu}^{c},
\label{eq:JSAA}
\\
J_{ASA}
&=&
\epsilon_{abc}
A_{ud}^{a;\mu}
S_{us}^{b}
A_{ds,\mu}^{c},
\label{eq:JASA}
\\
J_{AAS}
&=&
\epsilon_{abc}
A_{ud}^{a;\mu}
A_{us,\mu}^{b}
S_{ds}^{c}.
\label{eq:JAAS}
\end{eqnarray}
In each current, the Lorentz indices of the two axial-vector
diquarks are contracted to form a Lorentz scalar. Within the
$S$-wave assignments introduced above, these currents carry
$J^P=0^+$. Among them, $J_{SAA}$ is a pure $I=0$ current. In contrast,
$J_{ASA}$ and $J_{AAS}$, as written above, have $I_3=0$ but are
not pure isospin eigenstates. We nevertheless retain them as
auxiliary interpolating currents in order to examine how different
arrangements of scalar- and axial-vector-diquark correlations affect
the resulting QCD sum rules. Their couplings to scalar states are parametrized as
\begin{equation}
\langle 0|J_i|X_i(q)\rangle
=
f_i,
\qquad
i=SAA,\,ASA,\,AAS,
\label{eq:JSAA_couplings}
\end{equation}
where $|X_i(q)\rangle$ denotes a scalar state coupled to the
corresponding current and $f_i$ is the associated coupling constant.

We finally consider the configuration composed of three axial-vector
diquarks. Their Lorentz indices can be coupled through the totally
antisymmetric tensor to construct the Lorentz-vector current
\begin{equation}
J_{AAA}^{\mu}
=
\epsilon_{abc}
\epsilon^{\mu\nu\rho\sigma}
A_{ud,\nu}^{a}
A_{us,\rho}^{b}
A_{ds,\sigma}^{c}.
\label{eq:JAAA}
\end{equation}
As written above, this current also has $I_3=0$ but is not a pure
$I=0$ eigenstate. It is included as an additional probe of the
three-axial-vector-diquark correlation. Although $J_{AAA}^{\mu}$ carries
one Lorentz index, its longitudinal component can couple to a
$J^P=0^+$ state. We define this coupling by
\begin{equation}
\langle 0|J_{AAA}^{\mu}|X_{AAA}(q)\rangle
=
f_{AAA}\,q^{\mu},
\label{eq:JAAA_coupling}
\end{equation}
where $|X_{AAA}(q)\rangle$ denotes a $0^+$ state coupled to
$J_{AAA}^{\mu}$ and $f_{AAA}$ is the corresponding coupling
constant.

Together, the currents $J_{SSS}$, $J_{SAA}$, $J_{ASA}$, $J_{AAS}$,
and $J_{AAA}^{\mu}$ provide complementary probes of different
scalar- and axial-vector-diquark correlations in the $uuddss$
six-quark system. In particular, $J_{SSS}$ is an $SU(3)_{\rm flavor}$-singlet current
and hence necessarily carries $I=0$. The current $J_{SAA}$ is also a pure $I=0$ current, whereas
$J_{ASA}$, $J_{AAS}$, and $J_{AAA}^{\mu}$ are not pure isospin
eigenstates. The latter currents
are included to explore the sensitivity of the QCD sum rules to
different internal spin and flavor couplings, and are therefore not
interpreted as interpolators for a definite isoscalar $H$-dibaryon state.

\subsection{Baryon--Baryon Current}
\label{subsec:baryon}

We next construct a baryon--baryon interpolating current for the
$uuddss$ system. As discussed in Sec.~\ref{subsec:color}, we consider
the $\mathbf{1}_c\otimes\mathbf{1}_c$ color configuration, in which
the six quarks are organized into two color-singlet three-quark
clusters. For the isoscalar baryon--baryon sector considered here, we
choose the $\Lambda\Lambda$ configuration as a representative
baryon--baryon structure.

To construct the corresponding current, we first consider a local
interpolating current for the $\Lambda$ baryon. Local three-quark
baryon currents and their QCD sum rules have been extensively studied
in the literature~\cite{Ioffe:1981kw,Chung:1981cc,Espriu:1983hu,Chen:2008qv,Chen:2012ex}.
A local $SU(3)_{\rm flavor}$-octet $\Lambda$ current can be written as
\begin{equation}
\begin{aligned}
\eta_{\Lambda}^{\rm full}
=
\frac{1}{\sqrt{3}}\,
&\epsilon_{abc}
\Big[
2\left(d_a^T C s_b\right)\gamma_5 u_c
-2\left(u_a^T C s_b\right)\gamma_5 d_c
\\
&-4\left(u_a^T C d_b\right)\gamma_5 s_c
-2\left(d_a^T C\gamma_5 s_b\right)u_c
\\
&+2\left(u_a^T C\gamma_5 s_b\right)d_c
+4\left(u_a^T C\gamma_5 d_b\right)s_c
\Big].
\end{aligned}
\label{eq:Lambda_current_full}
\end{equation}
This current carries $I=0$ and has the appropriate quantum numbers
to couple to the spin-$1/2$ $\Lambda$ baryon.

To simplify the OPE calculation, in the present analysis we retain
only the explicit $(ud)s$ component of
Eq.~\eqref{eq:Lambda_current_full}, in which the $u$ and $d$ quarks
form the diquark pair while the strange quark acts as the third
quark. Absorbing the common overall normalization into the definition
of the current, we use
\begin{equation}
\eta_{\Lambda}
=
\epsilon_{abc}
\left[
\left(u_a^T C\gamma_5 d_b\right)s_c
-
\left(u_a^T C d_b\right)\gamma_5 s_c
\right].
\label{eq:Lambda_current}
\end{equation}
Both terms contain an isoscalar $ud$ pair, and hence
$\eta_{\Lambda}$ carries $I=0$ and has the appropriate flavor quantum
numbers to couple to the $\Lambda$ baryon.
Eq.~\eqref{eq:Lambda_current} should therefore be understood as
a restricted $(ud)s$ component of the full
$SU(3)_{\rm flavor}$-octet current in
Eq.~\eqref{eq:Lambda_current_full}, rather than as the complete
octet current.

The corresponding local $\Lambda\Lambda$ interpolating current is
constructed as
\begin{equation}
J_{\Lambda\Lambda}
=
\eta_{\Lambda}^{T} C\gamma_5\eta_{\Lambda}.
\label{eq:JLambdaLambda}
\end{equation}
The two spin-$1/2$ baryonic currents are coupled in the spin-singlet
channel. Within the $S$-wave assignment adopted here, the resulting
current carries $I=0$ and $J^P=0^+$. Its coupling to the corresponding
scalar state $|X_{\Lambda\Lambda}(q)\rangle$ is defined by
\begin{equation}
\langle 0|J_{\Lambda\Lambda}|X_{\Lambda\Lambda}(q)\rangle
=
f_{\Lambda\Lambda},
\label{eq:JLambdaLambda_coupling}
\end{equation}
where $f_{\Lambda\Lambda}$ denotes the corresponding coupling
constant.

For the $uuddss$ system, the $I=0$ baryon--baryon sector can in
general also involve other channels, such as $N\Xi$ and
$\Sigma\Sigma$. More complete $SU(3)_{\rm flavor}$-singlet
baryon--baryon constructions for the $H$ dibaryon have been
investigated in previous QCD sum-rule studies
~\cite{Larin:1985yt,Kodama:1994np,Zhang:2025qmg}.
In the present work, we restrict ourselves to the
$\Lambda\Lambda$ current defined in Eq.~\eqref{eq:JLambdaLambda}
and use it as a representative baryon--baryon benchmark for
comparison with the diquark--diquark--diquark currents constructed
in the previous subsection.

\section{QCD Sum Rule Analysis}
\label{sec:sumrule}

In this section, we develop the QCD sum rules for the $uuddss$
interpolating currents constructed in Sec.~\ref{sec:current}. To
illustrate the procedure, we first focus on the
$SU(3)_{\rm flavor}$-singlet current $J_{SSS}$ defined in
Eq.~\eqref{eq:JSSS}. We introduce the two-point correlation function
\begin{equation}
\Pi_{SSS}(q^2)
=
i\int d^4x\,e^{iq\cdot x}
\langle 0|
{\bf T}
\left[
J_{SSS}(x)J_{SSS}^{\dagger}(0)
\right]
|0\rangle .
\label{eq:PiSSS}
\end{equation}
As defined in Eq.~\eqref{eq:JSSS_coupling}, its coupling to a
$J^P=0^+$ state $|X_{SSS}(q)\rangle$ is parametrized by
\begin{equation}
\langle 0|J_{SSS}|X_{SSS}(q)\rangle
=
f_{SSS}.
\end{equation}
On the hadronic side, isolating the lowest-lying state gives
\begin{equation}
\Pi_{SSS}^{\rm phen}(q^2)
=
\frac{f_{SSS}^2}
{M_{SSS}^2-q^2}
+
\cdots ,
\label{eq:PiSSS_phen}
\end{equation}
where $M_{SSS}$ is the mass of the lowest state coupled to
$J_{SSS}$, and the ellipsis denotes higher resonances and continuum
states.

The hadronic correlation function can equivalently be expressed
through a dispersion relation,
\begin{equation}
\Pi_{SSS}^{\rm phen}(q^2)
=
\int_{s_{\rm min}}^{\infty}
\frac{\rho_{SSS}^{\rm phen}(s)}
{s-q^2-i\epsilon}\,ds
+
\mbox{subtractions},
\label{eq:dispersion_SSS}
\end{equation}
where $s_{\rm min}$ denotes the lower limit of the dispersion integral. 
We adopt the
usual pole-plus-continuum parametrization for the hadronic spectral
density,
\begin{equation}
\rho_{SSS}^{\rm phen}(s)
=
f_{SSS}^2\,
\delta(s-M_{SSS}^2)
+
\rho_{SSS}^{\rm higher}(s),
\label{eq:rhoSSS_phen}
\end{equation}
where $\rho_{SSS}^{\rm higher}(s)$ accounts for higher resonances and
continuum states.

At large spacelike momentum, the correlation function can be evaluated
at the quark-gluon level using the operator product expansion (OPE).
The short-distance contributions are contained in the Wilson
coefficients, whereas the nonperturbative QCD dynamics is encoded in
vacuum condensates of local operators. The corresponding spectral
density is defined by
\begin{equation}
\rho_{SSS}^{\rm OPE}(s)
=
\frac{1}{\pi}
{\rm Im}\,
\Pi_{SSS}^{\rm OPE}(s).
\label{eq:rhoSSS_OPE}
\end{equation}

To connect the hadronic and OPE descriptions, we invoke quark-hadron
duality and approximate the higher-state contribution by the OPE
spectral density above an effective continuum threshold $s_0$. We
then perform the Borel transformation, which removes the subtraction
terms and suppresses contributions from higher-energy states.
After continuum subtraction, the resulting sum rule for $J_{SSS}$
is
\begin{eqnarray}
{\cal L}_{SSS}(s_0,M_B^2)
&\equiv&
\int_{s_{\rm min}}^{s_0}
ds\,
e^{-s/M_B^2}
\rho_{SSS}^{\rm OPE}(s)
\nonumber\\
&=&
f_{SSS}^2
e^{-M_{SSS}^2/M_B^2}.
\label{eq:sumrule_SSS}
\end{eqnarray}
Here $M_B$ denotes the Borel mass, $s_0$ is the continuum
threshold, and ${\cal L}_{SSS}(s_0,M_B^2)$ represents the
continuum-subtracted zeroth Borel moment.
The hadron mass can be extracted by taking the derivative
of the sum rule with respect to $-1/M_B^2$ and dividing it by the
original sum rule,
\begin{equation}
M_{SSS}^2
=
\frac{
\dfrac{\partial}{\partial(-1/M_B^2)}
{\cal L}_{SSS}(s_0,M_B^2)
}{
{\cal L}_{SSS}(s_0,M_B^2)
}.
\label{eq:mass_SSS}
\end{equation}

The same procedure can be applied to the other currents,
\begin{equation}
J_{SAA},\quad
J_{ASA},\quad
J_{AAS},\quad
J_{AAA}^{\mu},\quad
J_{\Lambda\Lambda}.
\end{equation}
The explicit OPE expressions for the Borel moments of all six
interpolating currents are collected in Appendix~\ref{app:OPE}.
For all these correlation functions, the OPE is carried out up to
operators of dimension sixteen, without including radiative
corrections to the Wilson coefficients. Besides the perturbative
contribution, we retain the strange-quark mass, the light- and
strange-quark condensates, the gluon condensate, the quark-gluon
mixed condensates, and the higher-dimensional products of these
condensates that contribute up to $D=16$.
The higher-dimensional multiquark condensates
are estimated under the vacuum-saturation
approximation, introducing additional systematic
uncertainties associated with condensate factorization.
For the Lorentz-vector current $J_{AAA}^{\mu}$, we use the
longitudinal component that couples to the $J^P=0^+$ channel, as
specified in Sec.~\ref{sec:current}. Accordingly,
${\cal L}_{AAA}^{\rm OPE}(s_0,M_B^2)$ in
Appendix~\ref{app:OPE} denotes the Borel moment
of this component.

A common set of QCD input parameters and a uniform OPE truncation
scheme are employed for all the currents, allowing their sum-rule
behaviors to be compared consistently. In the next section, we
determine the working Borel windows and continuum thresholds by
examining the OPE convergence and pole contribution, and then present
the resulting mass estimates.

\section{Numerical Analysis}
\label{sec:numerical}

We now perform the numerical analyses based on the QCD sum rules
derived in Sec.~\ref{sec:sumrule}. Since the present six-quark
correlation functions contain condensate contributions of rather high
dimensions, the numerical behavior of the sum rules can depend
sensitively on the truncation of the OPE. We therefore examine the
successive truncations at
\begin{equation}
D=12,\quad D=14,\quad D=16,
\end{equation}
rather than assuming that the inclusion of higher-dimensional terms
necessarily improves the sum rule. Such a comparison allows us to
examine the truncation dependence of the results and to assess the
limitations of the present OPE treatment.

For a given OPE truncation up to dimension $D$, we first examine
the positivity of the continuum-subtracted Borel moment
${\cal L}(s_0,M_B^2)$. On the phenomenological side, the
lowest-lying pole contributes positively to this quantity.
Therefore, if the truncated OPE yields a sizable negative
Borel moment in the parameter region relevant to the sum-rule
analysis, we regard this as an indication that the corresponding
truncation is not suitable for the sum-rule extraction in this
region. Only after this basic consistency check do we proceed
to examine the Borel-window criteria.

The extracted mass depends on two external parameters: the Borel
mass $M_B$ and the continuum threshold $s_0$. Their working regions
are constrained mainly by the relative contribution of the
highest-dimensional OPE term and the pole contribution.
The continuum threshold $s_0$ is varied together with
$M_B^2$, and an acceptable sum-rule result should exhibit a
sufficiently stable mass prediction within the allowed parameter
region. We therefore determine the working ranges of $M_B^2$ and
$s_0$ by considering simultaneously the positivity of the Borel
moment, the relative contribution of the highest-dimensional
OPE term, the pole contribution, and the stability of the
extracted mass. The dependence on the OPE truncation is
examined separately to assess the reliability of the results.

For the numerical calculation, we use the following QCD input
parameters at the renormalization scale $\mu=2$~GeV
\cite{pdg,Ovchinnikov:1988gk,Yang:1993bp,Ellis:1996xc,
Ioffe:2002be,Jamin:2002ev,Gimenez:2005nt,Narison:2011xe,
Narison:2018dcr}:
\begin{eqnarray}
\nonumber
m_s
&=&
93^{+11}_{-5}~{\rm MeV} ,
\\
\nonumber
\langle\bar qq\rangle
&=&
-(0.240\pm0.010)^3~{\rm GeV}^3 ,
\\
\langle\bar ss\rangle
&=&
(0.8\pm0.1)\times\langle\bar qq\rangle ,
\label{eq:condensate}
\\
\nonumber
\langle g_s\bar q\sigma G q\rangle
&=&
-(0.8\pm0.2)\times\langle\bar qq\rangle~{\rm GeV}^2 ,
\\
\nonumber
\langle g_s\bar s\sigma G s\rangle
&=&
-(0.8\pm0.2)\times\langle\bar ss\rangle~{\rm GeV}^2 ,
\\
\nonumber
\langle\alpha_s GG\rangle
&=&
(6.35\pm0.35)\times10^{-2}~{\rm GeV}^4 .
\end{eqnarray}
We neglect the $u$- and $d$-quark masses, while retaining the
strange-quark mass and the $SU(3)$-flavor-breaking effects associated
with the strange-quark and mixed condensates.

In the following, we first present a detailed numerical analysis
of the $SU(3)_{\rm flavor}$-singlet current $J_{SSS}$, which contains three
scalar diquarks. We then compare its sum-rule behavior with that of
the other pure isoscalar current $J_{SAA}$, which contains one
scalar and two axial-vector diquarks, and briefly examine the
remaining auxiliary diquark currents. The baryon--baryon current
$J_{\Lambda\Lambda}$ is analyzed separately.

\subsection{Analysis of the $J_{SSS}$ Current}
\label{subsec:numerical_SSS}

\begin{figure}[hbtp]
\centering
\includegraphics[width=1.0\columnwidth]{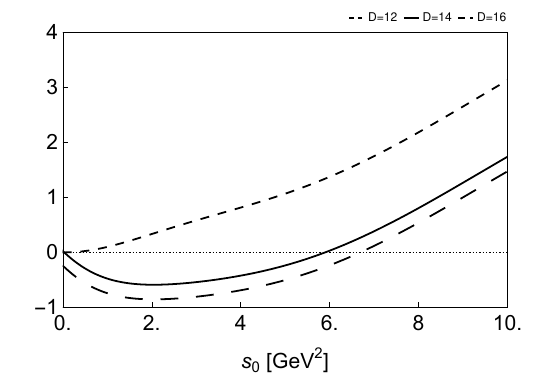}
\caption{
Continuum-subtracted Borel moments
${\cal L}_{SSS}(s_0,M_B^2)$ for the $J_{SSS}$ current
as functions of the continuum threshold $s_0$,
with $M_B^2=1.82~{\rm GeV}^2$.
The three curves correspond to the OPE truncations at
$D=12$ (short-dashed), $D=14$ (solid), and
$D=16$ (long-dashed).
The $D=14$ and $D=16$ results develop sizable
negative regions in the parameter range relevant to
the present analysis.
}
\label{fig:Pi_SSS}
\end{figure}

We begin with the $SU(3)_{\rm flavor}$-singlet current $J_{SSS}$,
which consists of three scalar diquarks. Since its OPE contains
condensate contributions of rather high dimensions, we first examine
the dependence of the Borel-transformed correlation function on
the OPE truncation. Fig.~\ref{fig:Pi_SSS} shows the continuum-subtracted
Borel moments ${\cal L}_{SSS}(s_0,M_B^2)$ as functions of $s_0$, with
$M_B^2=1.82~{\rm GeV}^2$, for the OPE truncated at
$D=12$, $14$, and $16$.
The $D=12$ result remains positive in the parameter
region relevant to the present analysis. In contrast, the
$D=14$ and $D=16$ results develop sizable negative
regions. Since the phenomenological pole contribution to the
Borel moment is positive, such negative values indicate that
the corresponding truncations are not suitable for the
sum-rule extraction in these parameter regions.

\begin{figure}[hbtp]
\centering
\includegraphics[width=1.00\columnwidth]{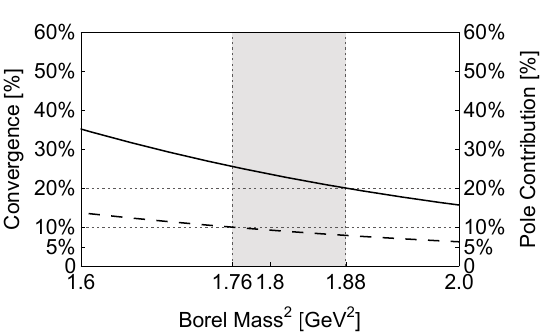}
\caption{
The convergence ratio CVG (dashed) and the pole contribution
PC (solid) as functions of the Borel mass $M_B^2$ for the
$J_{SSS}$ current with the OPE truncated at $D=12$, taking
$s_0=6.0~{\rm GeV}^2$. The shaded region denotes the
working Borel window determined by the adopted criteria.
}
\label{fig:CVGPC_SSS_D12}
\end{figure}

\begin{figure*}[htbp]
\centering
\includegraphics[width=0.46\textwidth]{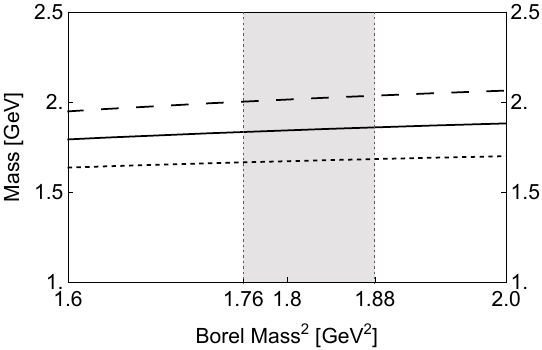}
\hfill
\includegraphics[width=0.46\textwidth]{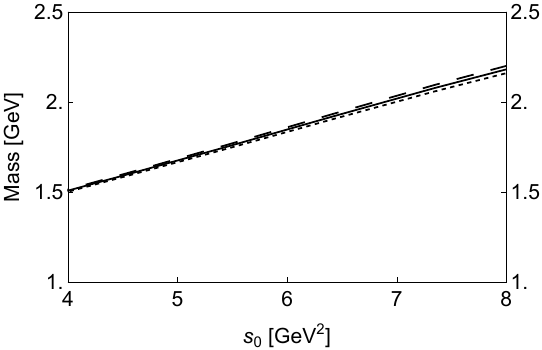}
\caption{
Mass extracted from the $J_{SSS}$ sum rule with the OPE
truncated at $D=12$.
Left: dependence on the Borel mass $M_B^2$ for
$s_0=5.0$ (short-dashed), $6.0$ (solid), and $7.0~{\rm GeV}^2$ (long-dashed).
Right: dependence on $s_0$ for
$M_B^2=1.76$ (short-dashed), $1.82$ (solid), and $1.88~{\rm GeV}^2$ (long-dashed).
}
\label{fig:mass_SSS_D12}
\end{figure*}

The positivity of the Borel moment alone is not sufficient
to establish an acceptable sum rule. We further examine
the relative contribution of the highest-dimensional
OPE term and the pole contribution, focusing on
the OPE truncated at $D=12$.
To examine the OPE convergence, we define the relative
contribution of the highest-dimensional term as
\begin{equation}
{\rm CVG}
=
\left|
\frac{
{\cal L}_{SSS}^{(12)}(\infty,M_B^2)
}{
{\cal L}_{SSS}^{(\leq12)}(\infty,M_B^2)
}
\right|,
\label{eq:cvg_SSS_D12}
\end{equation}
where ${\cal L}_{SSS}^{(12)}$ denotes the complete Borel
contribution of dimension twelve, while
${\cal L}_{SSS}^{(\leq12)}$ denotes the full OPE retained
through dimension twelve. The pole contribution is defined as
\begin{equation}
{\rm PC}
=
\frac{
{\cal L}_{SSS}^{(\leq12)}(s_0,M_B^2)
}{
{\cal L}_{SSS}^{(\leq12)}(\infty,M_B^2)
}.
\label{eq:pc_SSS_D12}
\end{equation}
We impose the following convergence and pole-contribution
criteria:
\begin{equation}
{\rm CVG}\leq10\%,
\qquad
{\rm PC}\geq20\%.
\label{eq:criteria_SSS_D12}
\end{equation}
These conditions are adopted to constrain the contribution
from the highest-dimensional OPE term and to determine
the working Borel window.
For high-dimensional six-quark currents, requiring a larger
pole contribution would favor smaller Borel masses, where
the OPE convergence becomes less reliable. We therefore
adopt ${\rm PC}\geq20\%$ as a practical compromise between
the pole contribution and OPE convergence~\cite{Chen:2014vha,Wan:2019ake,Chen:2019vdh,Ahmadi:2025vjv}.
The dependences of CVG and PC on the Borel mass are shown in
Fig.~\ref{fig:CVGPC_SSS_D12}. For the representative
continuum threshold $s_0=6.0~{\rm GeV}^2$, the adopted
criteria lead to the working Borel window
\begin{equation}
1.76~{\rm GeV}^2
\leq
M_B^2
\leq
1.88~{\rm GeV}^2.
\label{eq:Borel_SSS_D12}
\end{equation}

We then investigate the dependence of the extracted mass
on the two sum-rule parameters $M_B^2$ and $s_0$.
The results are shown in Fig.~\ref{fig:mass_SSS_D12}.
The extracted mass exhibits relatively stable behavior
with respect to $M_B^2$ within the working Borel window.
We further vary the continuum threshold in the range
\begin{equation}
5.0~{\rm GeV}^2
\leq
s_0
\leq
7.0~{\rm GeV}^2
\label{eq:s0_SSS_D12}
\end{equation}
to examine its influence on the extracted mass.
Within the parameter regions considered, we obtain
\begin{equation}
M_{SSS}^{D=12}
=
1.85^{+0.19}_{-0.20}~{\rm GeV},
\label{eq:mass_SSS_D12}
\end{equation}
where the central value is evaluated at
$M_B^2=1.82~{\rm GeV}^2$ and $s_0=6.0~{\rm GeV}^2$.
The uncertainties arise from variations of $M_B^2$ and $s_0$
within their working ranges, as well as from the uncertainties
in the QCD input parameters, within the $D=12$ truncation.

The mass extracted from the $D=12$ sum rule is compatible
with the phenomenologically motivated low-mass region.
We emphasize that this mass region is not used to define
the positivity, convergence, or pole-contribution criteria.
The present analysis therefore suggests that the
$SU(3)_{\rm flavor}$-singlet current $J_{SSS}$ can
accommodate the proposed low-mass $uuddss$ scenario
within the $D=12$ truncation. Nevertheless, the behavior
of the higher-dimensional OPE contributions indicates
a significant truncation dependence, which is not
included in the quoted mass uncertainty.

\subsection{Analysis of the Other Diquark Currents}
\label{subsec:numerical_other}

We next investigate the other four diquark--diquark--diquark
currents constructed in Sec.~\ref{sec:current},
\begin{equation}
J_{SAA},\qquad
J_{ASA},\qquad
J_{AAS},\qquad
J_{AAA}^{\mu}.
\end{equation}
Among them, $J_{SAA}$ is a pure isoscalar current, whereas
$J_{ASA}$, $J_{AAS}$, and $J_{AAA}^{\mu}$ carry $I_3=0$
but are not pure isospin eigenstates.

\begin{figure}[htbp]
\centering
\includegraphics[width=1.0\columnwidth]{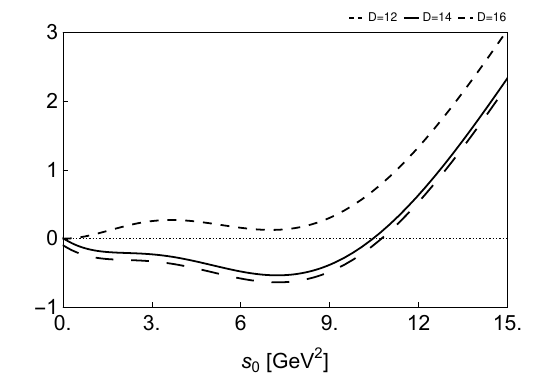}
\caption{
Continuum-subtracted Borel moments
${\cal L}_{SAA}(s_0,M_B^2)$ for the $J_{SAA}$ current
as functions of the continuum threshold $s_0$,
with $M_B^2=2.57~{\rm GeV}^2$.
The three curves correspond to the OPE truncations at
$D=12$ (short-dashed), $D=14$ (solid), and
$D=16$ (long-dashed).
The $D=14$ and $D=16$ results develop sizable
negative regions in the parameter range relevant to
the present analysis.
}
\label{fig:Pi_SAA}
\end{figure}

We first examine the $J_{SAA}$ current, which contains one scalar
and two axial-vector diquarks. Fig.~\ref{fig:Pi_SAA} shows its
continuum-subtracted Borel moments ${\cal L}_{SAA}(s_0,M_B^2)$
as functions of $s_0$, with $M_B^2=2.57~{\rm GeV}^2$,
for the OPE truncated at $D=12$, $14$, and $16$.
The results at $D=14$ and $D=16$ develop sizable
negative regions in the parameter range relevant to the
present analysis, indicating that these truncations are not
suitable for the sum-rule extraction in these regions.
In contrast, the $D=12$ Borel moment remains positive
in the relevant parameter region and therefore warrants
a separate examination of the sum-rule criteria.

We further investigate the $D=12$ truncation by examining
the relative contribution of the highest-dimensional OPE term,
the pole contribution, and the extracted mass.
Within the parameter region examined, we obtain
\begin{equation}
M_{SAA}^{D=12}>3.3~{\rm GeV},
\label{eq:mass_SAA_D12}
\end{equation}
considerably higher than the phenomenologically motivated
low-mass region. Consequently, the $J_{SAA}$ current does
not favor the proposed low-mass $uuddss$ scenario
within the $D=12$ truncation.

We have also examined the remaining currents $J_{ASA}$,
$J_{AAS}$, and $J_{AAA}^{\mu}$ using the same OPE truncation
schemes. Their numerical behaviors are generally similar
to that of $J_{SAA}$, with sizable negative Borel moments
appearing at certain truncation orders. At $D=12$, the
extracted masses satisfy
\begin{equation}
\begin{aligned}
M_{ASA}^{D=12} &> 3.3~{\rm GeV},\\
M_{AAS}^{D=12} &> 3.3~{\rm GeV},\\
M_{AAA}^{D=12} &> 3.9~{\rm GeV}.
\end{aligned}
\label{eq:mass_other_D12}
\end{equation}
These results likewise do not favor the proposed low-mass
$uuddss$ scenario within the $D=12$ truncation.

\subsection{Analysis of the Baryon--Baryon Current}
\label{subsec:numerical_BB}

We finally investigate the baryon--baryon current
$J_{\Lambda\Lambda}$ constructed in Sec.~\ref{sec:current},
which consists of two local $\Lambda$-type baryon currents.
This current provides a complementary description of the
$uuddss$ system in terms of color-singlet baryon clusters,
allowing us to compare its sum-rule behavior with that of
the diquark--diquark--diquark currents.

\begin{figure}[hbtp]
\centering
\includegraphics[width=1.0\columnwidth]{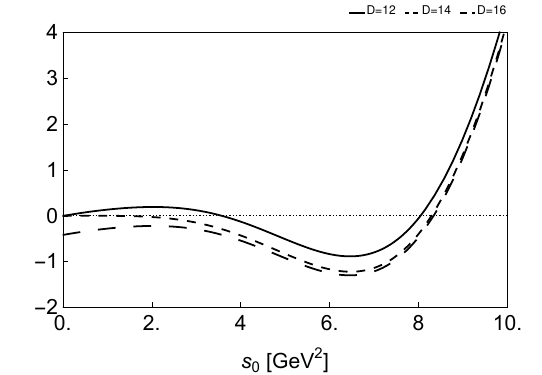}
\caption{
Continuum-subtracted Borel moments
${\cal L}_{\Lambda\Lambda}(s_0,M_B^2)$ for the
$J_{\Lambda\Lambda}$ current as functions of the
continuum threshold $s_0$, with
$M_B^2=2.50~{\rm GeV}^2$.
The three curves correspond to the OPE truncations at
$D=12$ (short-dashed), $D=14$ (solid), and
$D=16$ (long-dashed).
All three results develop sizable negative regions
in the parameter range considered.
}
\label{fig:Pi_LL}
\end{figure}

We first examine the dependence of its continuum-subtracted
Borel moment ${\cal L}_{\Lambda\Lambda}(s_0,M_B^2)$ on the
OPE truncation. Fig.~\ref{fig:Pi_LL} shows the results as
functions of $s_0$, with $M_B^2=2.50~{\rm GeV}^2$, for the
OPE truncated at $D=12$, $14$, and $16$.
All three truncations develop sizable negative Borel moments
over intermediate ranges of $s_0$, indicating that the
corresponding sum rules are not suitable for mass extraction
in these regions. In particular, negative Borel moments
already appear at $D=12$, while the differences among
the three curves reveal a significant dependence on the
higher-dimensional OPE contributions.

We further examine the $D=12$ truncation by considering
the positivity of the Borel moment, the relative contribution
of the highest-dimensional OPE term, the pole contribution,
and the stability of the extracted mass.
However, a sufficiently reliable working Borel window
cannot be established within the parameter regions examined.
We therefore refrain from extracting a quantitative mass
prediction from the $J_{\Lambda\Lambda}$ sum rule.

\subsection{Comparison with Previous Theoretical Studies}
\label{subsec:numerical_comparison}

The $H$ dibaryon has been investigated using QCD sum rules
in several previous studies. Larin et al.~\cite{Larin:1985yt}
examined six candidate local interpolating currents and
obtained mass estimates of approximately $2.0$ and $2.4$~GeV
without and with the strange-quark mass, respectively.
They concluded that their results could not determine
whether the $H$ dibaryon lies above or below the
$\Lambda\Lambda$ threshold.
A particularly relevant calculation was performed by
Azizi, Agaev, and Sundu~\cite{Azizi:2019xla}, who studied
the scalar $uuddss$ hexaquark using three scalar diquarks.
Their interpolating current is algebraically identical
to our $J_{SSS}$ current.
They obtained masses of
$1.180^{+0.040}_{-0.026}$~GeV at $\mu=2$~GeV and
$1.239^{+0.042}_{-0.028}$~GeV at $\mu=1$~GeV,
both lower than our $D=12$ estimate.
The two analyses employ different OPE truncations,
continuum thresholds, and Borel-window prescriptions.
In particular, Ref.~\cite{Azizi:2019xla} uses
$s_0=2.5$--$2.9~{\rm GeV}^2$, whereas the present
$D=12$ analysis considers $s_0=5.0$--$7.0~{\rm GeV}^2$.

The baryon--baryon currents employed in previous studies
also deserve attention.
Kodama, Oka, and Hatsuda~\cite{Kodama:1994np}
constructed an $SU(3)_{\rm flavor}$-singlet current
from two flavor-octet baryon currents.
They obtained a mass near $1.88$~GeV in the
flavor-symmetric limit, with the continuum threshold
calibrated against the di-nucleon channel, and approximately
$2.19$~GeV after including $SU(3)$-flavor-breaking effects,
the latter lying about $40$~MeV below the
$\Lambda\Lambda$ threshold.
Paganis et al.~\cite{Paganis:1997rf} subsequently
extended the QCD sum-rule analysis to
flavor-nonsinglet dibaryon channels.
The flavor-singlet current constructed in
Ref.~\cite{Kodama:1994np} involves the complete
octet--octet combination, including the
$\Lambda\Lambda$, $N\Xi$, and $\Sigma\Sigma$ channels.
It therefore differs from the reduced
$J_{\Lambda\Lambda}$ current considered in the
present work, and a more systematic analysis of
baryon--baryon interpolating currents is still needed.

It is also worth noting that a $uuddss$ state with a
mass near $1.85$~GeV would lie approximately $380$~MeV
below the physical $\Lambda\Lambda$ threshold.
Such a deeply bound state is difficult to accommodate
within many existing hadronic descriptions.
Chiral effective field theory studies have illustrated
the sensitivity of the $H$-dibaryon binding to the
underlying interactions and flavor-symmetry
breaking~\cite{Haidenbauer:2011ah}.
More recently, Hu and Guo~\cite{Hu:2026hdb}
investigated coupled-channel baryon--baryon
interactions within an $SU(6)$-constrained effective
field theory. They obtained two bound-state poles
below the $\Lambda\Lambda$ threshold, with central
masses near $2.13$ and $2.22$~GeV for one of their
input prescriptions, although the shallower pole
is sensitive to the cutoff variation.
These results illustrate the important role of
coupled-channel dynamics.
Lattice QCD studies have found weak binding in the
flavor-symmetric limit at heavier-than-physical
quark masses or weak low-energy $\Lambda\Lambda$
attraction near the physical quark masses
~\cite{Green:2021qol,HALQCD:2019wsz}.
Although these results do not exclude a deeply
bound compact $uuddss$ state, they indicate that
a binding energy as large as $380$~MeV remains
challenging to explain within existing hadronic
approaches.

\section{Summary and Outlook}
\label{sec:summary}

In this work, we have investigated the $J^P=0^+$
$uuddss$ hexaquark system using QCD sum rules.
We have considered five local diquark--diquark--diquark
interpolating currents involving different combinations
of scalar and axial-vector diquarks, together with
a baryon--baryon current composed of two local
$\Lambda$-type baryon currents. The operator product
expansions have been calculated up to dimension sixteen,
and the numerical analyses have been performed by
examining the positivity of the Borel moments,
the contributions of higher-dimensional condensates,
the pole contributions, and the stability of the
extracted masses.

Among the interpolating currents considered,
the $SU(3)_{\rm flavor}$-singlet current $J_{SSS}$,
which consists of three scalar diquarks, exhibits
particularly interesting behavior. With the OPE
truncated at $D=12$, we obtain
\begin{equation}
M_{SSS}^{D=12}
=
1.85^{+0.19}_{-0.20}~{\rm GeV},
\end{equation}
compatible with the phenomenologically motivated
low-mass $uuddss$ scenario. In contrast, the other
four diquark currents containing axial-vector diquarks
yield considerably higher mass estimates within the
parameter regions examined at the same truncation level.
For the baryon--baryon current $J_{\Lambda\Lambda}$,
a sufficiently reliable working Borel window cannot
be established, and we therefore refrain from
providing a quantitative mass prediction.

The numerical results also exhibit a significant
dependence on the OPE truncation. In particular,
the inclusion of dimension-fourteen and dimension-sixteen
contributions leads to negative Borel moments in parts
of the relevant parameter region. Consequently,
the mass obtained at $D=12$ should be regarded as
a conditional estimate within the adopted truncation
scheme, with additional systematic uncertainties
associated with higher-dimensional OPE contributions
that are not included in the quoted uncertainty.

An interesting feature of the $J_{SSS}$ current is
that scalar diquarks with $J^P=0^+$ and color
$\bar{\mathbf{3}}_C$ belong to the flavor
$\bar{\mathbf{3}}_F$ representation.
For the three light quark flavors $u$, $d$, and $s$,
the independent scalar-diquark combinations are
$[ud]$, $[ds]$, and $[su]$.
Since local scalar-diquark operators commute with
one another, their antisymmetric color contraction
vanishes whenever two of them are identical.
Consequently, up to normalization, the only nonvanishing
local color-singlet current constructed from three
such scalar diquarks is
\begin{equation}
J_{SSS}
=
\epsilon_{abc}S_{ud}^{a}S_{ds}^{b}S_{su}^{c},
\end{equation}
which is an $SU(3)_{\rm flavor}$ singlet with
the quark content $uuddss$.
In particular, analogous all-scalar-diquark currents
vanish for the $uuuddd$ and $uuudds$ systems.
In contrast, axial-vector diquarks with $J^P=1^+$
and color $\bar{\mathbf{3}}_C$ belong to the
flavor $\mathbf{6}_F$ representation,
allowing a wider variety of six-quark flavor
configurations, including $uuuddd$ and $uuudds$.
Together with our QCD sum-rule results at $D=12$,
this observation motivates further investigation
of whether the special flavor and spin correlations
associated with three scalar diquarks can favor
a low-mass $uuddss$ state.

A $uuddss$ state with a mass near $1.85$~GeV
would lie approximately $380$~MeV below the
$\Lambda\Lambda$ threshold, corresponding to a
much deeper binding than that typically obtained
from conventional baryon--baryon interactions.
This suggests that, if such a state exists,
its low mass may be associated with specific
multiquark correlations rather than ordinary
baryon--baryon binding alone.

The possible existence of such a deeply bound state
is subject to important phenomenological constraints.
Besides the stability of ordinary nuclei, studies
of deeply bound dibaryons in nearby mass regions
have identified potentially stringent constraints
from neutron stars and supernovae~\cite{McDermott:2018ofd}.
Its interpretation as a significant component of
dark matter also faces nontrivial cosmological
constraints, particularly those associated with
thermal production and interactions with ordinary
baryonic matter~\cite{Kolb:2018bxv,Moore:2024mot}.
Alternative nonthermal production mechanisms have
also been investigated~\cite{Moore:2025zyg}.
These constraints depend on additional transition
matrix elements, effective interactions, and
cosmological assumptions that cannot be determined
from the present two-point QCD sum rules alone.
Accordingly, our results should be interpreted as
an investigation of the possible low-mass six-quark
spectrum, rather than as an establishment of a
phenomenologically viable dark matter candidate.

Nevertheless, the unique flavor structure of
the local all-scalar-diquark current does not
establish the uniqueness of the physical state,
and different interpolating currents may couple
to the same state. In particular, a more systematic
analysis of baryon--baryon currents, including
their flavor structures and coupled-channel
contributions, is needed to investigate possible
molecular components~\cite{Chen:2021xlu}. Further improvements in
the treatment of higher-dimensional OPE contributions
will also be essential for determining whether
the low-mass solution survives a more reliable
QCD sum-rule analysis.

\section*{Acknowledgments}

This work is supported by
the National Natural Science Foundation of China under Grant No.~12675096,
the Jiangsu Provincial Double-Innovation Program under Grant No.~JSSCRC2021488, 
and
the Fundamental Research Funds for the Central Universities.

\appendix

\begin{widetext}
\section{OPE Expressions for the Borel Moments}
\label{app:OPE}

In this appendix, we present the explicit OPE expressions
for the continuum-subtracted Borel moments of the six
interpolating currents considered in the numerical analysis:
\begin{eqnarray}
&& {\cal L}_{SSS}^{\rm OPE}(s_0,M_B^2) = \hspace*{0pt plus 1filll} \label{app:rhoSSS} \\
&&  \int_{4m_s^2}^{s_0} \left(  +\frac{s^7}{61656268800\,\pi^{10}} -\frac{17\,m_s^2\,\,s^6}{9909043200\,\pi^{10}} + s^5\,\left(  -\frac{m_s\,\langle\bar qq\rangle}{7372800\,\pi^{8}} +\frac{7\,m_s\,\langle\bar ss\rangle}{176947200\,\pi^{8}} +\frac{\langle\alpha_sGG\rangle}{419430400\,\pi^{10}} \right) \right. \nonumber\\
&& \left. \qquad + s^4\,\left(  -\frac{271\,\langle\alpha_sGG\rangle\,m_s^2}{3019898880\,\pi^{10}} +\frac{7\,\langle\bar qq\rangle^{2}}{2211840\,\pi^{6}} -\frac{19\,m_s\,\langle g_s\bar q\sigma Gq\rangle}{20971520\,\pi^{8}} +\frac{\langle\bar qq\rangle\,\langle\bar ss\rangle}{184320\,\pi^{6}} +\frac{\langle\bar ss\rangle^{2}}{4423680\,\pi^{6}} -\frac{m_s\,\langle g_s\bar s\sigma Gs\rangle}{37748736\,\pi^{8}} \right) \right. \nonumber\\
&& \left. \qquad + s^3\,\left(  -\frac{53\,\langle\alpha_sGG\rangle\,m_s\,\langle\bar qq\rangle}{18874368\,\pi^{8}} -\frac{7\,m_s^2\,\langle\bar qq\rangle^{2}}{110592\,\pi^{6}} +\frac{181\,\langle\bar qq\rangle\,\langle g_s\bar q\sigma Gq\rangle}{7077888\,\pi^{6}} +\frac{55\,\langle\alpha_sGG\rangle\,m_s\,\langle\bar ss\rangle}{56623104\,\pi^{8}} -\frac{m_s^2\,\langle\bar qq\rangle\,\langle\bar ss\rangle}{6144\,\pi^{6}} \right.\right. \nonumber\\
&& \left.\left. \qquad +\frac{19\,\langle g_s\bar q\sigma Gq\rangle\,\langle\bar ss\rangle}{786432\,\pi^{6}} +\frac{m_s^2\,\langle\bar ss\rangle^{2}}{98304\,\pi^{6}} +\frac{19\,\langle\bar qq\rangle\,\langle g_s\bar s\sigma Gs\rangle}{786432\,\pi^{6}} +\frac{5\,\langle\bar ss\rangle\,\langle g_s\bar s\sigma Gs\rangle}{7077888\,\pi^{6}} \right) + s^2\,\left(  +\frac{193\,\langle\alpha_sGG\rangle\,\langle\bar qq\rangle^{2}}{7077888\,\pi^{6}} \right.\right. \nonumber\\
&& \left.\left. \qquad -\frac{m_s\,\langle\bar qq\rangle^{3}}{576\,\pi^{4}} -\frac{53\,\langle\alpha_sGG\rangle\,m_s\,\langle g_s\bar q\sigma Gq\rangle}{6291456\,\pi^{8}} -\frac{181\,m_s^2\,\langle\bar qq\rangle\,\langle g_s\bar q\sigma Gq\rangle}{589824\,\pi^{6}} +\frac{23\,\langle g_s\bar q\sigma Gq\rangle^{2}}{786432\,\pi^{6}} +\frac{53\,\langle\alpha_sGG\rangle\,\langle\bar qq\rangle\,\langle\bar ss\rangle}{1179648\,\pi^{6}} \right.\right. \nonumber\\
&& \left.\left. \qquad -\frac{7\,m_s\,\langle\bar qq\rangle^{2}\,\langle\bar ss\rangle}{6912\,\pi^{4}} -\frac{91\,m_s^2\,\langle g_s\bar q\sigma Gq\rangle\,\langle\bar ss\rangle}{196608\,\pi^{6}} +\frac{17\,\langle\alpha_sGG\rangle\,\langle\bar ss\rangle^{2}}{7077888\,\pi^{6}} +\frac{m_s\,\langle\bar qq\rangle\,\langle\bar ss\rangle^{2}}{1152\,\pi^{4}} -\frac{17\,\langle\alpha_sGG\rangle\,m_s\,\langle g_s\bar s\sigma Gs\rangle}{18874368\,\pi^{8}} \right.\right. \nonumber\\
&& \left.\left. \qquad -\frac{15\,m_s^2\,\langle\bar qq\rangle\,\langle g_s\bar s\sigma Gs\rangle}{65536\,\pi^{6}} +\frac{25\,\langle g_s\bar q\sigma Gq\rangle\,\langle g_s\bar s\sigma Gs\rangle}{393216\,\pi^{6}} -\frac{\langle g_s\bar s\sigma Gs\rangle^{2}}{786432\,\pi^{6}} \right) + s\,\left(  -\frac{65\,\langle\alpha_sGG\rangle\,m_s^2\,\langle\bar qq\rangle^{2}}{393216\,\pi^{6}} +\frac{\langle\bar qq\rangle^{4}}{1728\,\pi^{2}} \right.\right. \nonumber\\
&& \left.\left. \qquad +\frac{193\,\langle\alpha_sGG\rangle\,\langle\bar qq\rangle\,\langle g_s\bar q\sigma Gq\rangle}{2359296\,\pi^{6}} -\frac{509\,m_s\,\langle\bar qq\rangle^{2}\,\langle g_s\bar q\sigma Gq\rangle}{73728\,\pi^{4}} -\frac{23\,m_s^2\,\langle g_s\bar q\sigma Gq\rangle^{2}}{131072\,\pi^{6}} -\frac{25\,\langle\alpha_sGG\rangle\,m_s^2\,\langle\bar qq\rangle\,\langle\bar ss\rangle}{196608\,\pi^{6}} \right.\right. \nonumber\\
&& \left.\left. \qquad +\frac{\langle\bar qq\rangle^{3}\,\langle\bar ss\rangle}{72\,\pi^{2}} +\frac{53\,\langle\alpha_sGG\rangle\,\langle g_s\bar q\sigma Gq\rangle\,\langle\bar ss\rangle}{786432\,\pi^{6}} -\frac{533\,m_s\,\langle\bar qq\rangle\,\langle g_s\bar q\sigma Gq\rangle\,\langle\bar ss\rangle}{147456\,\pi^{4}} +\frac{19\,\langle\alpha_sGG\rangle\,m_s^2\,\langle\bar ss\rangle^{2}}{1572864\,\pi^{6}} +\frac{7\,\langle\bar qq\rangle^{2}\,\langle\bar ss\rangle^{2}}{864\,\pi^{2}} \right.\right. \nonumber\\
&& \left.\left. \qquad +\frac{91\,m_s\,\langle g_s\bar q\sigma Gq\rangle\,\langle\bar ss\rangle^{2}}{73728\,\pi^{4}} +\frac{53\,\langle\alpha_sGG\rangle\,\langle\bar qq\rangle\,\langle g_s\bar s\sigma Gs\rangle}{786432\,\pi^{6}} -\frac{167\,m_s\,\langle\bar qq\rangle^{2}\,\langle g_s\bar s\sigma Gs\rangle}{73728\,\pi^{4}} -\frac{125\,m_s^2\,\langle g_s\bar q\sigma Gq\rangle\,\langle g_s\bar s\sigma Gs\rangle}{393216\,\pi^{6}} \right.\right. \nonumber\\
&& \left.\left. \qquad +\frac{17\,\langle\alpha_sGG\rangle\,\langle\bar ss\rangle\,\langle g_s\bar s\sigma Gs\rangle}{2359296\,\pi^{6}} +\frac{15\,m_s\,\langle\bar qq\rangle\,\langle\bar ss\rangle\,\langle g_s\bar s\sigma Gs\rangle}{16384\,\pi^{4}} \right) + \left(  -\frac{29\,\langle\alpha_sGG\rangle\,m_s\,\langle\bar qq\rangle^{3}}{73728\,\pi^{4}} +\frac{7\,m_s^2\,\langle\bar qq\rangle^{4}}{864\,\pi^{2}} \right.\right. \nonumber\\
&& \left.\left. \qquad -\frac{65\,\langle\alpha_sGG\rangle\,m_s^2\,\langle\bar qq\rangle\,\langle g_s\bar q\sigma Gq\rangle}{393216\,\pi^{6}} +\frac{7\,\langle\bar qq\rangle^{3}\,\langle g_s\bar q\sigma Gq\rangle}{9216\,\pi^{2}} +\frac{193\,\langle\alpha_sGG\rangle\,\langle g_s\bar q\sigma Gq\rangle^{2}}{9437184\,\pi^{6}} -\frac{221\,m_s\,\langle\bar qq\rangle\,\langle g_s\bar q\sigma Gq\rangle^{2}}{73728\,\pi^{4}} \right.\right. \nonumber\\
&& \left.\left. \qquad +\frac{\langle\alpha_sGG\rangle\,m_s\,\langle\bar qq\rangle^{2}\,\langle\bar ss\rangle}{36864\,\pi^{4}} -\frac{m_s^2\,\langle\bar qq\rangle^{3}\,\langle\bar ss\rangle}{24\,\pi^{2}} -\frac{25\,\langle\alpha_sGG\rangle\,m_s^2\,\langle g_s\bar q\sigma Gq\rangle\,\langle\bar ss\rangle}{393216\,\pi^{6}} +\frac{509\,\langle\bar qq\rangle^{2}\,\langle g_s\bar q\sigma Gq\rangle\,\langle\bar ss\rangle}{27648\,\pi^{2}} \right.\right. \nonumber\\
&& \left.\left. \qquad -\frac{103\,m_s\,\langle g_s\bar q\sigma Gq\rangle^{2}\,\langle\bar ss\rangle}{98304\,\pi^{4}} +\frac{25\,\langle\alpha_sGG\rangle\,m_s\,\langle\bar qq\rangle\,\langle\bar ss\rangle^{2}}{221184\,\pi^{4}} +\frac{7\,m_s^2\,\langle\bar qq\rangle^{2}\,\langle\bar ss\rangle^{2}}{1728\,\pi^{2}} +\frac{119\,\langle\bar qq\rangle\,\langle g_s\bar q\sigma Gq\rangle\,\langle\bar ss\rangle^{2}}{13824\,\pi^{2}} \right.\right. \nonumber\\
&& \left.\left. \qquad -\frac{25\,\langle\alpha_sGG\rangle\,m_s^2\,\langle\bar qq\rangle\,\langle g_s\bar s\sigma Gs\rangle}{589824\,\pi^{6}} +\frac{217\,\langle\bar qq\rangle^{3}\,\langle g_s\bar s\sigma Gs\rangle}{27648\,\pi^{2}} +\frac{53\,\langle\alpha_sGG\rangle\,\langle g_s\bar q\sigma Gq\rangle\,\langle g_s\bar s\sigma Gs\rangle}{1572864\,\pi^{6}} \right.\right. \nonumber\\
&& \left.\left. \qquad -\frac{181\,m_s\,\langle\bar qq\rangle\,\langle g_s\bar q\sigma Gq\rangle\,\langle g_s\bar s\sigma Gs\rangle}{73728\,\pi^{4}} +\frac{167\,\langle\bar qq\rangle^{2}\,\langle\bar ss\rangle\,\langle g_s\bar s\sigma Gs\rangle}{27648\,\pi^{2}} +\frac{125\,m_s\,\langle g_s\bar q\sigma Gq\rangle\,\langle\bar ss\rangle\,\langle g_s\bar s\sigma Gs\rangle}{294912\,\pi^{4}} \right.\right. \nonumber\\
&& \left.\left. \qquad +\frac{17\,\langle\alpha_sGG\rangle\,\langle g_s\bar s\sigma Gs\rangle^{2}}{9437184\,\pi^{6}} \right) \right)\,e^{-s/M_B^2}\,ds - \left(  -\frac{5\,\langle\alpha_sGG\rangle\,\langle\bar qq\rangle^{4}}{110592\,\pi^{2}} -\frac{29\,\langle\alpha_sGG\rangle\,m_s\,\langle\bar qq\rangle^{2}\,\langle g_s\bar q\sigma Gq\rangle}{98304\,\pi^{4}} \right. \nonumber\\
&& \left. \qquad +\frac{235\,m_s^2\,\langle\bar qq\rangle^{3}\,\langle g_s\bar q\sigma Gq\rangle}{27648\,\pi^{2}} -\frac{65\,\langle\alpha_sGG\rangle\,m_s^2\,\langle g_s\bar q\sigma Gq\rangle^{2}}{3145728\,\pi^{6}} +\frac{5\,\langle\bar qq\rangle^{2}\,\langle g_s\bar q\sigma Gq\rangle^{2}}{36864\,\pi^{2}} -\frac{125\,m_s\,\langle g_s\bar q\sigma Gq\rangle^{3}}{589824\,\pi^{4}} \right. \nonumber\\
&& \left. \qquad +\frac{29\,\langle\alpha_sGG\rangle\,\langle\bar qq\rangle^{3}\,\langle\bar ss\rangle}{55296\,\pi^{2}} -\frac{17}{648}\,m_s\,\langle\bar qq\rangle^{4}\,\langle\bar ss\rangle +\frac{\langle\alpha_sGG\rangle\,m_s\,\langle\bar qq\rangle\,\langle g_s\bar q\sigma Gq\rangle\,\langle\bar ss\rangle}{73728\,\pi^{4}} -\frac{283\,m_s^2\,\langle\bar qq\rangle^{2}\,\langle g_s\bar q\sigma Gq\rangle\,\langle\bar ss\rangle}{9216\,\pi^{2}} \right. \nonumber\\
&& \left. \qquad +\frac{221\,\langle\bar qq\rangle\,\langle g_s\bar q\sigma Gq\rangle^{2}\,\langle\bar ss\rangle}{55296\,\pi^{2}} +\frac{19\,\langle\alpha_sGG\rangle\,\langle\bar qq\rangle^{2}\,\langle\bar ss\rangle^{2}}{110592\,\pi^{2}} +\frac{1}{54}\,m_s\,\langle\bar qq\rangle^{3}\,\langle\bar ss\rangle^{2} +\frac{25\,\langle\alpha_sGG\rangle\,m_s\,\langle g_s\bar q\sigma Gq\rangle\,\langle\bar ss\rangle^{2}}{884736\,\pi^{4}} \right. \nonumber\\
&& \left. \qquad +\frac{119\,m_s^2\,\langle\bar qq\rangle\,\langle g_s\bar q\sigma Gq\rangle\,\langle\bar ss\rangle^{2}}{55296\,\pi^{2}} +\frac{7\,\langle g_s\bar q\sigma Gq\rangle^{2}\,\langle\bar ss\rangle^{2}}{6144\,\pi^{2}} -\frac{19\,\langle\alpha_sGG\rangle\,m_s\,\langle\bar qq\rangle^{2}\,\langle g_s\bar s\sigma Gs\rangle}{294912\,\pi^{4}} -\frac{181\,m_s^2\,\langle\bar qq\rangle^{3}\,\langle g_s\bar s\sigma Gs\rangle}{27648\,\pi^{2}} \right. \nonumber\\
&& \left. \qquad -\frac{25\,\langle\alpha_sGG\rangle\,m_s^2\,\langle g_s\bar q\sigma Gq\rangle\,\langle g_s\bar s\sigma Gs\rangle}{2359296\,\pi^{6}} +\frac{73\,\langle\bar qq\rangle^{2}\,\langle g_s\bar q\sigma Gq\rangle\,\langle g_s\bar s\sigma Gs\rangle}{13824\,\pi^{2}} -\frac{65\,m_s\,\langle g_s\bar q\sigma Gq\rangle^{2}\,\langle g_s\bar s\sigma Gs\rangle}{196608\,\pi^{4}} \right. \nonumber\\
&& \left. \qquad +\frac{25\,\langle\alpha_sGG\rangle\,m_s\,\langle\bar qq\rangle\,\langle\bar ss\rangle\,\langle g_s\bar s\sigma Gs\rangle}{884736\,\pi^{4}} +\frac{181\,\langle\bar qq\rangle\,\langle g_s\bar q\sigma Gq\rangle\,\langle\bar ss\rangle\,\langle g_s\bar s\sigma Gs\rangle}{55296\,\pi^{2}} +\frac{55\,\langle\bar qq\rangle^{2}\,\langle g_s\bar s\sigma Gs\rangle^{2}}{110592\,\pi^{2}} \right), \nonumber
\\[5mm]
&& {\cal L}_{SAA}^{\rm OPE}(s_0,M_B^2) = \hspace*{0pt plus 1filll} \label{app:rhoSAA} \\
&&  \int_{4m_s^2}^{s_0} \left(  +\frac{s^7}{15414067200\,\pi^{10}} -\frac{31\,m_s^2\,\,s^6}{4954521600\,\pi^{10}} + s^5\,\left(  -\frac{m_s\,\langle\bar qq\rangle}{3686400\,\pi^{8}} +\frac{m_s\,\langle\bar ss\rangle}{11059200\,\pi^{8}} +\frac{\langle\alpha_sGG\rangle}{471859200\,\pi^{10}} \right) \right. \nonumber\\
&& \left. \qquad + s^4\,\left(  -\frac{17\,\langle\alpha_sGG\rangle\,m_s^2}{150994944\,\pi^{10}} +\frac{5\,\langle\bar qq\rangle^{2}}{442368\,\pi^{6}} -\frac{11\,m_s\,\langle g_s\bar q\sigma Gq\rangle}{5242880\,\pi^{8}} +\frac{\langle\bar qq\rangle\,\langle\bar ss\rangle}{92160\,\pi^{6}} +\frac{\langle\bar ss\rangle^{2}}{442368\,\pi^{6}} -\frac{7\,m_s\,\langle g_s\bar s\sigma Gs\rangle}{9437184\,\pi^{8}} \right) \right. \nonumber\\
&& \left. \qquad + s^3\,\left(  +\frac{5\,\langle\alpha_sGG\rangle\,m_s\,\langle\bar qq\rangle}{1572864\,\pi^{8}} -\frac{25\,m_s^2\,\langle\bar qq\rangle^{2}}{110592\,\pi^{6}} +\frac{67\,\langle\bar qq\rangle\,\langle g_s\bar q\sigma Gq\rangle}{442368\,\pi^{6}} -\frac{7\,\langle\alpha_sGG\rangle\,m_s\,\langle\bar ss\rangle}{7077888\,\pi^{8}} -\frac{m_s^2\,\langle\bar qq\rangle\,\langle\bar ss\rangle}{3072\,\pi^{6}} \right.\right. \nonumber\\
&& \left.\left. \qquad +\frac{11\,\langle g_s\bar q\sigma Gq\rangle\,\langle\bar ss\rangle}{196608\,\pi^{6}} +\frac{m_s^2\,\langle\bar ss\rangle^{2}}{24576\,\pi^{6}} +\frac{49\,\langle\bar qq\rangle\,\langle g_s\bar s\sigma Gs\rangle}{589824\,\pi^{6}} +\frac{35\,\langle\bar ss\rangle\,\langle g_s\bar s\sigma Gs\rangle}{1769472\,\pi^{6}} \right) + s^2\,\left(  -\frac{\langle\alpha_sGG\rangle\,\langle\bar qq\rangle^{2}}{221184\,\pi^{6}} \right.\right. \nonumber\\
&& \left.\left. \qquad -\frac{m_s\,\langle\bar qq\rangle^{3}}{288\,\pi^{4}} +\frac{5\,\langle\alpha_sGG\rangle\,m_s\,\langle g_s\bar q\sigma Gq\rangle}{524288\,\pi^{8}} -\frac{67\,m_s^2\,\langle\bar qq\rangle\,\langle g_s\bar q\sigma Gq\rangle}{36864\,\pi^{6}} +\frac{7\,\langle g_s\bar q\sigma Gq\rangle^{2}}{24576\,\pi^{6}} -\frac{5\,\langle\alpha_sGG\rangle\,\langle\bar qq\rangle\,\langle\bar ss\rangle}{98304\,\pi^{6}} \right.\right. \nonumber\\
&& \left.\left. \qquad -\frac{25\,m_s\,\langle\bar qq\rangle^{2}\,\langle\bar ss\rangle}{6912\,\pi^{4}} -\frac{7\,m_s^2\,\langle g_s\bar q\sigma Gq\rangle\,\langle\bar ss\rangle}{8192\,\pi^{6}} +\frac{47\,\langle\alpha_sGG\rangle\,\langle\bar ss\rangle^{2}}{1769472\,\pi^{6}} +\frac{m_s\,\langle\bar qq\rangle\,\langle\bar ss\rangle^{2}}{576\,\pi^{4}} -\frac{47\,\langle\alpha_sGG\rangle\,m_s\,\langle g_s\bar s\sigma Gs\rangle}{4718592\,\pi^{8}} \right.\right. \nonumber\\
&& \left.\left. \qquad -\frac{41\,m_s^2\,\langle\bar qq\rangle\,\langle g_s\bar s\sigma Gs\rangle}{49152\,\pi^{6}} +\frac{25\,\langle g_s\bar q\sigma Gq\rangle\,\langle g_s\bar s\sigma Gs\rangle}{98304\,\pi^{6}} +\frac{5\,\langle g_s\bar s\sigma Gs\rangle^{2}}{196608\,\pi^{6}} \right) + s\,\left(  -\frac{7\,\langle\alpha_sGG\rangle\,m_s^2\,\langle\bar qq\rangle^{2}}{147456\,\pi^{6}} +\frac{5\,\langle\bar qq\rangle^{4}}{864\,\pi^{2}} \right.\right. \nonumber\\
&& \left.\left. \qquad -\frac{\langle\alpha_sGG\rangle\,\langle\bar qq\rangle\,\langle g_s\bar q\sigma Gq\rangle}{73728\,\pi^{6}} -\frac{101\,m_s\,\langle\bar qq\rangle^{2}\,\langle g_s\bar q\sigma Gq\rangle}{6144\,\pi^{4}} -\frac{7\,m_s^2\,\langle g_s\bar q\sigma Gq\rangle^{2}}{4096\,\pi^{6}} -\frac{\langle\alpha_sGG\rangle\,m_s^2\,\langle\bar qq\rangle\,\langle\bar ss\rangle}{8192\,\pi^{6}} \right.\right. \nonumber\\
&& \left.\left. \qquad +\frac{\langle\bar qq\rangle^{3}\,\langle\bar ss\rangle}{36\,\pi^{2}} -\frac{5\,\langle\alpha_sGG\rangle\,\langle g_s\bar q\sigma Gq\rangle\,\langle\bar ss\rangle}{65536\,\pi^{6}} -\frac{485\,m_s\,\langle\bar qq\rangle\,\langle g_s\bar q\sigma Gq\rangle\,\langle\bar ss\rangle}{36864\,\pi^{4}} +\frac{\langle\alpha_sGG\rangle\,m_s^2\,\langle\bar ss\rangle^{2}}{18432\,\pi^{6}} +\frac{25\,\langle\bar qq\rangle^{2}\,\langle\bar ss\rangle^{2}}{864\,\pi^{2}} \right.\right. \nonumber\\
&& \left.\left. \qquad +\frac{7\,m_s\,\langle g_s\bar q\sigma Gq\rangle\,\langle\bar ss\rangle^{2}}{3072\,\pi^{4}} -\frac{5\,\langle\alpha_sGG\rangle\,\langle\bar qq\rangle\,\langle g_s\bar s\sigma Gs\rangle}{65536\,\pi^{6}} -\frac{217\,m_s\,\langle\bar qq\rangle^{2}\,\langle g_s\bar s\sigma Gs\rangle}{18432\,\pi^{4}} -\frac{37\,m_s^2\,\langle g_s\bar q\sigma Gq\rangle\,\langle g_s\bar s\sigma Gs\rangle}{32768\,\pi^{6}} \right.\right. \nonumber\\
&& \left.\left. \qquad +\frac{47\,\langle\alpha_sGG\rangle\,\langle\bar ss\rangle\,\langle g_s\bar s\sigma Gs\rangle}{589824\,\pi^{6}} +\frac{41\,m_s\,\langle\bar qq\rangle\,\langle\bar ss\rangle\,\langle g_s\bar s\sigma Gs\rangle}{12288\,\pi^{4}} \right) + \left(  +\frac{5\,\langle\alpha_sGG\rangle\,m_s\,\langle\bar qq\rangle^{3}}{6144\,\pi^{4}} +\frac{m_s^2\,\langle\bar qq\rangle^{4}}{54\,\pi^{2}} \right.\right. \nonumber\\
&& \left.\left. \qquad -\frac{7\,\langle\alpha_sGG\rangle\,m_s^2\,\langle\bar qq\rangle\,\langle g_s\bar q\sigma Gq\rangle}{147456\,\pi^{6}} +\frac{25\,\langle\bar qq\rangle^{3}\,\langle g_s\bar q\sigma Gq\rangle}{2304\,\pi^{2}} -\frac{\langle\alpha_sGG\rangle\,\langle g_s\bar q\sigma Gq\rangle^{2}}{294912\,\pi^{6}} -\frac{53\,m_s\,\langle\bar qq\rangle\,\langle g_s\bar q\sigma Gq\rangle^{2}}{6144\,\pi^{4}} \right.\right. \nonumber\\
&& \left.\left. \qquad -\frac{83\,\langle\alpha_sGG\rangle\,m_s\,\langle\bar qq\rangle^{2}\,\langle\bar ss\rangle}{110592\,\pi^{4}} -\frac{m_s^2\,\langle\bar qq\rangle^{3}\,\langle\bar ss\rangle}{12\,\pi^{2}} -\frac{\langle\alpha_sGG\rangle\,m_s^2\,\langle g_s\bar q\sigma Gq\rangle\,\langle\bar ss\rangle}{16384\,\pi^{6}} +\frac{101\,\langle\bar qq\rangle^{2}\,\langle g_s\bar q\sigma Gq\rangle\,\langle\bar ss\rangle}{2304\,\pi^{2}} \right.\right. \nonumber\\
&& \left.\left. \qquad -\frac{95\,m_s\,\langle g_s\bar q\sigma Gq\rangle^{2}\,\langle\bar ss\rangle}{24576\,\pi^{4}} +\frac{\langle\alpha_sGG\rangle\,m_s\,\langle\bar qq\rangle\,\langle\bar ss\rangle^{2}}{9216\,\pi^{4}} +\frac{25\,m_s^2\,\langle\bar qq\rangle^{2}\,\langle\bar ss\rangle^{2}}{1728\,\pi^{2}} +\frac{251\,\langle\bar qq\rangle\,\langle g_s\bar q\sigma Gq\rangle\,\langle\bar ss\rangle^{2}}{6912\,\pi^{2}} \right.\right. \nonumber\\
&& \left.\left. \qquad -\frac{\langle\alpha_sGG\rangle\,m_s^2\,\langle\bar qq\rangle\,\langle g_s\bar s\sigma Gs\rangle}{24576\,\pi^{6}} +\frac{5\,\langle\bar qq\rangle^{3}\,\langle g_s\bar s\sigma Gs\rangle}{256\,\pi^{2}} -\frac{5\,\langle\alpha_sGG\rangle\,\langle g_s\bar q\sigma Gq\rangle\,\langle g_s\bar s\sigma Gs\rangle}{131072\,\pi^{6}} \right.\right. \nonumber\\
&& \left.\left. \qquad -\frac{67\,m_s\,\langle\bar qq\rangle\,\langle g_s\bar q\sigma Gq\rangle\,\langle g_s\bar s\sigma Gs\rangle}{4608\,\pi^{4}} +\frac{217\,\langle\bar qq\rangle^{2}\,\langle\bar ss\rangle\,\langle g_s\bar s\sigma Gs\rangle}{6912\,\pi^{2}} +\frac{37\,m_s\,\langle g_s\bar q\sigma Gq\rangle\,\langle\bar ss\rangle\,\langle g_s\bar s\sigma Gs\rangle}{24576\,\pi^{4}} \right.\right. \nonumber\\
&& \left.\left. \qquad +\frac{47\,\langle\alpha_sGG\rangle\,\langle g_s\bar s\sigma Gs\rangle^{2}}{2359296\,\pi^{6}} \right) \right)\,e^{-s/M_B^2}\,ds - \left(  -\frac{7\,\langle\alpha_sGG\rangle\,\langle\bar qq\rangle^{4}}{27648\,\pi^{2}} +\frac{5\,\langle\alpha_sGG\rangle\,m_s\,\langle\bar qq\rangle^{2}\,\langle g_s\bar q\sigma Gq\rangle}{8192\,\pi^{4}} \right. \nonumber\\
&& \left. \qquad +\frac{133\,m_s^2\,\langle\bar qq\rangle^{3}\,\langle g_s\bar q\sigma Gq\rangle}{6912\,\pi^{2}} -\frac{7\,\langle\alpha_sGG\rangle\,m_s^2\,\langle g_s\bar q\sigma Gq\rangle^{2}}{1179648\,\pi^{6}} +\frac{35\,\langle\bar qq\rangle^{2}\,\langle g_s\bar q\sigma Gq\rangle^{2}}{9216\,\pi^{2}} -\frac{37\,m_s\,\langle g_s\bar q\sigma Gq\rangle^{3}}{49152\,\pi^{4}} \right. \nonumber\\
&& \left. \qquad -\frac{5\,\langle\alpha_sGG\rangle\,\langle\bar qq\rangle^{3}\,\langle\bar ss\rangle}{4608\,\pi^{2}} -\frac{31}{324}\,m_s\,\langle\bar qq\rangle^{4}\,\langle\bar ss\rangle -\frac{83\,\langle\alpha_sGG\rangle\,m_s\,\langle\bar qq\rangle\,\langle g_s\bar q\sigma Gq\rangle\,\langle\bar ss\rangle}{221184\,\pi^{4}} -\frac{23\,m_s^2\,\langle\bar qq\rangle^{2}\,\langle g_s\bar q\sigma Gq\rangle\,\langle\bar ss\rangle}{384\,\pi^{2}} \right. \nonumber\\
&& \left. \qquad +\frac{53\,\langle\bar qq\rangle\,\langle g_s\bar q\sigma Gq\rangle^{2}\,\langle\bar ss\rangle}{4608\,\pi^{2}} +\frac{5\,\langle\alpha_sGG\rangle\,\langle\bar qq\rangle^{2}\,\langle\bar ss\rangle^{2}}{6912\,\pi^{2}} +\frac{1}{27}\,m_s\,\langle\bar qq\rangle^{3}\,\langle\bar ss\rangle^{2} +\frac{\langle\alpha_sGG\rangle\,m_s\,\langle g_s\bar q\sigma Gq\rangle\,\langle\bar ss\rangle^{2}}{36864\,\pi^{4}} \right. \nonumber\\
&& \left. \qquad +\frac{251\,m_s^2\,\langle\bar qq\rangle\,\langle g_s\bar q\sigma Gq\rangle\,\langle\bar ss\rangle^{2}}{27648\,\pi^{2}} +\frac{151\,\langle g_s\bar q\sigma Gq\rangle^{2}\,\langle\bar ss\rangle^{2}}{27648\,\pi^{2}} -\frac{5\,\langle\alpha_sGG\rangle\,m_s\,\langle\bar qq\rangle^{2}\,\langle g_s\bar s\sigma Gs\rangle}{18432\,\pi^{4}} -\frac{37\,m_s^2\,\langle\bar qq\rangle^{3}\,\langle g_s\bar s\sigma Gs\rangle}{2304\,\pi^{2}} \right. \nonumber\\
&& \left. \qquad -\frac{\langle\alpha_sGG\rangle\,m_s^2\,\langle g_s\bar q\sigma Gq\rangle\,\langle g_s\bar s\sigma Gs\rangle}{98304\,\pi^{6}} +\frac{35\,\langle\bar qq\rangle^{2}\,\langle g_s\bar q\sigma Gq\rangle\,\langle g_s\bar s\sigma Gs\rangle}{2304\,\pi^{2}} -\frac{319\,m_s\,\langle g_s\bar q\sigma Gq\rangle^{2}\,\langle g_s\bar s\sigma Gs\rangle}{147456\,\pi^{4}} \right. \nonumber\\
&& \left. \qquad +\frac{\langle\alpha_sGG\rangle\,m_s\,\langle\bar qq\rangle\,\langle\bar ss\rangle\,\langle g_s\bar s\sigma Gs\rangle}{36864\,\pi^{4}} +\frac{67\,\langle\bar qq\rangle\,\langle g_s\bar q\sigma Gq\rangle\,\langle\bar ss\rangle\,\langle g_s\bar s\sigma Gs\rangle}{3456\,\pi^{2}} +\frac{13\,\langle\bar qq\rangle^{2}\,\langle g_s\bar s\sigma Gs\rangle^{2}}{3072\,\pi^{2}} \right), \nonumber
\\[5mm]
&& {\cal L}_{ASA}^{\rm OPE}(s_0,M_B^2) = {\cal L}_{AAS}^{\rm OPE}(s_0,M_B^2) = \hspace*{0pt plus 1filll} \label{app:rhoASA} \\
&&  \int_{4m_s^2}^{s_0} \left(  +\frac{s^7}{15414067200\,\pi^{10}} -\frac{71\,m_s^2\,\,s^6}{9909043200\,\pi^{10}} + s^5\,\left(  -\frac{m_s\,\langle\bar qq\rangle}{2457600\,\pi^{8}} +\frac{17\,m_s\,\langle\bar ss\rangle}{88473600\,\pi^{8}} +\frac{\langle\alpha_sGG\rangle}{471859200\,\pi^{10}} \right) \right. \nonumber\\
&& \left. \qquad + s^4\,\left(  -\frac{83\,\langle\alpha_sGG\rangle\,m_s^2}{1509949440\,\pi^{10}} +\frac{7\,\langle\bar qq\rangle^{2}}{884736\,\pi^{6}} -\frac{13\,m_s\,\langle g_s\bar q\sigma Gq\rangle}{3145728\,\pi^{8}} +\frac{\langle\bar qq\rangle\,\langle\bar ss\rangle}{61440\,\pi^{6}} +\frac{\langle\bar ss\rangle^{2}}{4423680\,\pi^{6}} -\frac{5\,m_s\,\langle g_s\bar s\sigma Gs\rangle}{18874368\,\pi^{8}} \right) \right. \nonumber\\
&& \left. \qquad + s^3\,\left(  +\frac{5\,\langle\alpha_sGG\rangle\,m_s\,\langle\bar qq\rangle}{4718592\,\pi^{8}} -\frac{35\,m_s^2\,\langle\bar qq\rangle^{2}}{221184\,\pi^{6}} +\frac{341\,\langle\bar qq\rangle\,\langle g_s\bar q\sigma Gq\rangle}{3538944\,\pi^{6}} +\frac{19\,\langle\alpha_sGG\rangle\,m_s\,\langle\bar ss\rangle}{14155776\,\pi^{8}} -\frac{m_s^2\,\langle\bar qq\rangle\,\langle\bar ss\rangle}{2048\,\pi^{6}} \right.\right. \nonumber\\
&& \left.\left. \qquad +\frac{65\,\langle g_s\bar q\sigma Gq\rangle\,\langle\bar ss\rangle}{589824\,\pi^{6}} +\frac{m_s^2\,\langle\bar ss\rangle^{2}}{24576\,\pi^{6}} +\frac{19\,\langle\bar qq\rangle\,\langle g_s\bar s\sigma Gs\rangle}{196608\,\pi^{6}} +\frac{25\,\langle\bar ss\rangle\,\langle g_s\bar s\sigma Gs\rangle}{3538944\,\pi^{6}} \right) + s^2\,\left(  -\frac{19\,\langle\alpha_sGG\rangle\,\langle\bar qq\rangle^{2}}{3538944\,\pi^{6}} \right.\right. \nonumber\\
&& \left.\left. \qquad -\frac{m_s\,\langle\bar qq\rangle^{3}}{192\,\pi^{4}} +\frac{5\,\langle\alpha_sGG\rangle\,m_s\,\langle g_s\bar q\sigma Gq\rangle}{1572864\,\pi^{8}} -\frac{341\,m_s^2\,\langle\bar qq\rangle\,\langle g_s\bar q\sigma Gq\rangle}{294912\,\pi^{6}} +\frac{67\,\langle g_s\bar q\sigma Gq\rangle^{2}}{393216\,\pi^{6}} -\frac{5\,\langle\alpha_sGG\rangle\,\langle\bar qq\rangle\,\langle\bar ss\rangle}{294912\,\pi^{6}} \right.\right. \nonumber\\
&& \left.\left. \qquad -\frac{35\,m_s\,\langle\bar qq\rangle^{2}\,\langle\bar ss\rangle}{13824\,\pi^{4}} -\frac{m_s^2\,\langle g_s\bar q\sigma Gq\rangle\,\langle\bar ss\rangle}{512\,\pi^{6}} -\frac{23\,\langle\alpha_sGG\rangle\,\langle\bar ss\rangle^{2}}{3538944\,\pi^{6}} +\frac{m_s\,\langle\bar qq\rangle\,\langle\bar ss\rangle^{2}}{384\,\pi^{4}} +\frac{23\,\langle\alpha_sGG\rangle\,m_s\,\langle g_s\bar s\sigma Gs\rangle}{9437184\,\pi^{8}} \right.\right. \nonumber\\
&& \left.\left. \qquad -\frac{71\,m_s^2\,\langle\bar qq\rangle\,\langle g_s\bar s\sigma Gs\rangle}{73728\,\pi^{6}} +\frac{37\,\langle g_s\bar q\sigma Gq\rangle\,\langle g_s\bar s\sigma Gs\rangle}{98304\,\pi^{6}} +\frac{7\,\langle g_s\bar s\sigma Gs\rangle^{2}}{393216\,\pi^{6}} \right) + s\,\left(  -\frac{19\,\langle\alpha_sGG\rangle\,m_s^2\,\langle\bar qq\rangle^{2}}{65536\,\pi^{6}} +\frac{\langle\bar qq\rangle^{4}}{1728\,\pi^{2}} \right.\right. \nonumber\\
&& \left.\left. \qquad -\frac{19\,\langle\alpha_sGG\rangle\,\langle\bar qq\rangle\,\langle g_s\bar q\sigma Gq\rangle}{1179648\,\pi^{6}} -\frac{239\,m_s\,\langle\bar qq\rangle^{2}\,\langle g_s\bar q\sigma Gq\rangle}{9216\,\pi^{4}} -\frac{67\,m_s^2\,\langle g_s\bar q\sigma Gq\rangle^{2}}{65536\,\pi^{6}} +\frac{53\,\langle\alpha_sGG\rangle\,m_s^2\,\langle\bar qq\rangle\,\langle\bar ss\rangle}{98304\,\pi^{6}} \right.\right. \nonumber\\
&& \left.\left. \qquad +\frac{\langle\bar qq\rangle^{3}\,\langle\bar ss\rangle}{24\,\pi^{2}} -\frac{5\,\langle\alpha_sGG\rangle\,\langle g_s\bar q\sigma Gq\rangle\,\langle\bar ss\rangle}{196608\,\pi^{6}} -\frac{691\,m_s\,\langle\bar qq\rangle\,\langle g_s\bar q\sigma Gq\rangle\,\langle\bar ss\rangle}{73728\,\pi^{4}} -\frac{17\,\langle\alpha_sGG\rangle\,m_s^2\,\langle\bar ss\rangle^{2}}{589824\,\pi^{6}} +\frac{35\,\langle\bar qq\rangle^{2}\,\langle\bar ss\rangle^{2}}{1728\,\pi^{2}} \right.\right. \nonumber\\
&& \left.\left. \qquad +\frac{m_s\,\langle g_s\bar q\sigma Gq\rangle\,\langle\bar ss\rangle^{2}}{192\,\pi^{4}} -\frac{5\,\langle\alpha_sGG\rangle\,\langle\bar qq\rangle\,\langle g_s\bar s\sigma Gs\rangle}{196608\,\pi^{6}} -\frac{277\,m_s\,\langle\bar qq\rangle^{2}\,\langle g_s\bar s\sigma Gs\rangle}{36864\,\pi^{4}} -\frac{95\,m_s^2\,\langle g_s\bar q\sigma Gq\rangle\,\langle g_s\bar s\sigma Gs\rangle}{49152\,\pi^{6}} \right.\right. \nonumber\\
&& \left.\left. \qquad -\frac{23\,\langle\alpha_sGG\rangle\,\langle\bar ss\rangle\,\langle g_s\bar s\sigma Gs\rangle}{1179648\,\pi^{6}} +\frac{71\,m_s\,\langle\bar qq\rangle\,\langle\bar ss\rangle\,\langle g_s\bar s\sigma Gs\rangle}{18432\,\pi^{4}} \right) + \left(  -\frac{7\,\langle\alpha_sGG\rangle\,m_s\,\langle\bar qq\rangle^{3}}{18432\,\pi^{4}} +\frac{17\,m_s^2\,\langle\bar qq\rangle^{4}}{432\,\pi^{2}} \right.\right. \nonumber\\
&& \left.\left. \qquad -\frac{19\,\langle\alpha_sGG\rangle\,m_s^2\,\langle\bar qq\rangle\,\langle g_s\bar q\sigma Gq\rangle}{65536\,\pi^{6}} +\frac{11\,\langle\bar qq\rangle^{3}\,\langle g_s\bar q\sigma Gq\rangle}{4608\,\pi^{2}} -\frac{19\,\langle\alpha_sGG\rangle\,\langle g_s\bar q\sigma Gq\rangle^{2}}{4718592\,\pi^{6}} -\frac{131\,m_s\,\langle\bar qq\rangle\,\langle g_s\bar q\sigma Gq\rangle^{2}}{9216\,\pi^{4}} \right.\right. \nonumber\\
&& \left.\left. \qquad +\frac{3\,\langle\alpha_sGG\rangle\,m_s\,\langle\bar qq\rangle^{2}\,\langle\bar ss\rangle}{2048\,\pi^{4}} -\frac{m_s^2\,\langle\bar qq\rangle^{3}\,\langle\bar ss\rangle}{8\,\pi^{2}} +\frac{53\,\langle\alpha_sGG\rangle\,m_s^2\,\langle g_s\bar q\sigma Gq\rangle\,\langle\bar ss\rangle}{196608\,\pi^{6}} +\frac{239\,\langle\bar qq\rangle^{2}\,\langle g_s\bar q\sigma Gq\rangle\,\langle\bar ss\rangle}{3456\,\pi^{2}} \right.\right. \nonumber\\
&& \left.\left. \qquad -\frac{137\,m_s\,\langle g_s\bar q\sigma Gq\rangle^{2}\,\langle\bar ss\rangle}{49152\,\pi^{4}} +\frac{53\,\langle\alpha_sGG\rangle\,m_s\,\langle\bar qq\rangle\,\langle\bar ss\rangle^{2}}{110592\,\pi^{4}} +\frac{35\,m_s^2\,\langle\bar qq\rangle^{2}\,\langle\bar ss\rangle^{2}}{3456\,\pi^{2}} +\frac{43\,\langle\bar qq\rangle\,\langle g_s\bar q\sigma Gq\rangle\,\langle\bar ss\rangle^{2}}{1728\,\pi^{2}} \right.\right. \nonumber\\
&& \left.\left. \qquad +\frac{53\,\langle\alpha_sGG\rangle\,m_s^2\,\langle\bar qq\rangle\,\langle g_s\bar s\sigma Gs\rangle}{294912\,\pi^{6}} +\frac{11\,\langle\bar qq\rangle^{3}\,\langle g_s\bar s\sigma Gs\rangle}{432\,\pi^{2}} -\frac{5\,\langle\alpha_sGG\rangle\,\langle g_s\bar q\sigma Gq\rangle\,\langle g_s\bar s\sigma Gs\rangle}{393216\,\pi^{6}} \right.\right. \nonumber\\
&& \left.\left. \qquad -\frac{341\,m_s\,\langle\bar qq\rangle\,\langle g_s\bar q\sigma Gq\rangle\,\langle g_s\bar s\sigma Gs\rangle}{36864\,\pi^{4}} +\frac{277\,\langle\bar qq\rangle^{2}\,\langle\bar ss\rangle\,\langle g_s\bar s\sigma Gs\rangle}{13824\,\pi^{2}} +\frac{95\,m_s\,\langle g_s\bar q\sigma Gq\rangle\,\langle\bar ss\rangle\,\langle g_s\bar s\sigma Gs\rangle}{36864\,\pi^{4}} \right.\right. \nonumber\\
&& \left.\left. \qquad -\frac{23\,\langle\alpha_sGG\rangle\,\langle g_s\bar s\sigma Gs\rangle^{2}}{4718592\,\pi^{6}} \right) \right)\,e^{-s/M_B^2}\,ds - \left(  -\frac{\langle\alpha_sGG\rangle\,\langle\bar qq\rangle^{4}}{6144\,\pi^{2}} -\frac{7\,\langle\alpha_sGG\rangle\,m_s\,\langle\bar qq\rangle^{2}\,\langle g_s\bar q\sigma Gq\rangle}{24576\,\pi^{4}} \right. \nonumber\\
&& \left. \qquad +\frac{527\,m_s^2\,\langle\bar qq\rangle^{3}\,\langle g_s\bar q\sigma Gq\rangle}{13824\,\pi^{2}} -\frac{19\,\langle\alpha_sGG\rangle\,m_s^2\,\langle g_s\bar q\sigma Gq\rangle^{2}}{524288\,\pi^{6}} +\frac{25\,\langle\bar qq\rangle^{2}\,\langle g_s\bar q\sigma Gq\rangle^{2}}{18432\,\pi^{2}} -\frac{95\,m_s\,\langle g_s\bar q\sigma Gq\rangle^{3}}{73728\,\pi^{4}} \right. \nonumber\\
&& \left. \qquad +\frac{7\,\langle\alpha_sGG\rangle\,\langle\bar qq\rangle^{3}\,\langle\bar ss\rangle}{13824\,\pi^{2}} -\frac{71}{648}\,m_s\,\langle\bar qq\rangle^{4}\,\langle\bar ss\rangle +\frac{3\,\langle\alpha_sGG\rangle\,m_s\,\langle\bar qq\rangle\,\langle g_s\bar q\sigma Gq\rangle\,\langle\bar ss\rangle}{4096\,\pi^{4}} -\frac{5\,m_s^2\,\langle\bar qq\rangle^{2}\,\langle g_s\bar q\sigma Gq\rangle\,\langle\bar ss\rangle}{48\,\pi^{2}} \right. \nonumber\\
&& \left. \qquad +\frac{131\,\langle\bar qq\rangle\,\langle g_s\bar q\sigma Gq\rangle^{2}\,\langle\bar ss\rangle}{6912\,\pi^{2}} +\frac{53\,\langle\alpha_sGG\rangle\,\langle\bar qq\rangle^{2}\,\langle\bar ss\rangle^{2}}{55296\,\pi^{2}} +\frac{1}{18}\,m_s\,\langle\bar qq\rangle^{3}\,\langle\bar ss\rangle^{2} -\frac{53\,\langle\alpha_sGG\rangle\,m_s\,\langle g_s\bar q\sigma Gq\rangle\,\langle\bar ss\rangle^{2}}{442368\,\pi^{4}} \right. \nonumber\\
&& \left. \qquad +\frac{43\,m_s^2\,\langle\bar qq\rangle\,\langle g_s\bar q\sigma Gq\rangle\,\langle\bar ss\rangle^{2}}{6912\,\pi^{2}} +\frac{17\,\langle g_s\bar q\sigma Gq\rangle^{2}\,\langle\bar ss\rangle^{2}}{4608\,\pi^{2}} +\frac{53\,\langle\alpha_sGG\rangle\,m_s\,\langle\bar qq\rangle^{2}\,\langle g_s\bar s\sigma Gs\rangle}{147456\,\pi^{4}} -\frac{49\,m_s^2\,\langle\bar qq\rangle^{3}\,\langle g_s\bar s\sigma Gs\rangle}{2304\,\pi^{2}} \right. \nonumber\\
&& \left. \qquad +\frac{53\,\langle\alpha_sGG\rangle\,m_s^2\,\langle g_s\bar q\sigma Gq\rangle\,\langle g_s\bar s\sigma Gs\rangle}{1179648\,\pi^{6}} +\frac{287\,\langle\bar qq\rangle^{2}\,\langle g_s\bar q\sigma Gq\rangle\,\langle g_s\bar s\sigma Gs\rangle}{13824\,\pi^{2}} -\frac{45\,m_s\,\langle g_s\bar q\sigma Gq\rangle^{2}\,\langle g_s\bar s\sigma Gs\rangle}{32768\,\pi^{4}} \right. \nonumber\\
&& \left. \qquad +\frac{53\,\langle\alpha_sGG\rangle\,m_s\,\langle\bar qq\rangle\,\langle\bar ss\rangle\,\langle g_s\bar s\sigma Gs\rangle}{442368\,\pi^{4}} +\frac{341\,\langle\bar qq\rangle\,\langle g_s\bar q\sigma Gq\rangle\,\langle\bar ss\rangle\,\langle g_s\bar s\sigma Gs\rangle}{27648\,\pi^{2}} +\frac{137\,\langle\bar qq\rangle^{2}\,\langle g_s\bar s\sigma Gs\rangle^{2}}{55296\,\pi^{2}} \right), \nonumber
\\[5mm]
&& {\cal L}_{AAA}^{\rm OPE}(s_0,M_B^2) = \hspace*{0pt plus 1filll} \label{app:rhoAAA} \\
&&  \int_{4m_s^2}^{s_0} \left(  +\frac{s^7}{13872660480\,\pi^{10}} -\frac{17\,m_s^2\,\,s^6}{2642411520\,\pi^{10}} + s^5\,\left(  -\frac{17\,m_s\,\langle\bar qq\rangle}{20643840\,\pi^{8}} +\frac{m_s\,\langle\bar ss\rangle}{14745600\,\pi^{8}} -\frac{29\,\langle\alpha_sGG\rangle}{3303014400\,\pi^{10}} \right) \right. \nonumber\\
&& \left. \qquad + s^4\,\left(  +\frac{311\,\langle\alpha_sGG\rangle\,m_s^2}{1132462080\,\pi^{10}} +\frac{91\,\langle\bar qq\rangle^{2}}{4423680\,\pi^{6}} -\frac{287\,m_s\,\langle g_s\bar q\sigma Gq\rangle}{35389440\,\pi^{8}} +\frac{23\,\langle\bar qq\rangle\,\langle\bar ss\rangle}{737280\,\pi^{6}} +\frac{11\,\langle\bar ss\rangle^{2}}{4423680\,\pi^{6}} -\frac{13\,m_s\,\langle g_s\bar s\sigma Gs\rangle}{11796480\,\pi^{8}} \right) \right. \nonumber\\
&& \left. \qquad + s^3\,\left(  +\frac{67\,\langle\alpha_sGG\rangle\,m_s\,\langle\bar qq\rangle}{5898240\,\pi^{8}} -\frac{73\,m_s^2\,\langle\bar qq\rangle^{2}}{184320\,\pi^{6}} +\frac{21\,\langle\bar qq\rangle\,\langle g_s\bar q\sigma Gq\rangle}{81920\,\pi^{6}} -\frac{23\,\langle\alpha_sGG\rangle\,m_s\,\langle\bar ss\rangle}{17694720\,\pi^{8}} -\frac{53\,m_s^2\,\langle\bar qq\rangle\,\langle\bar ss\rangle}{61440\,\pi^{6}} \right.\right. \nonumber\\
&& \left.\left. \qquad +\frac{49\,\langle g_s\bar q\sigma Gq\rangle\,\langle\bar ss\rangle}{245760\,\pi^{6}} +\frac{m_s^2\,\langle\bar ss\rangle^{2}}{36864\,\pi^{6}} +\frac{49\,\langle\bar qq\rangle\,\langle g_s\bar s\sigma Gs\rangle}{245760\,\pi^{6}} +\frac{7\,\langle\bar ss\rangle\,\langle g_s\bar s\sigma Gs\rangle}{245760\,\pi^{6}} \right) + s^2\,\left(  -\frac{211\,\langle\alpha_sGG\rangle\,\langle\bar qq\rangle^{2}}{1769472\,\pi^{6}} \right.\right. \nonumber\\
&& \left.\left. \qquad -\frac{43\,m_s\,\langle\bar qq\rangle^{3}}{4608\,\pi^{4}} +\frac{145\,\langle\alpha_sGG\rangle\,m_s\,\langle g_s\bar q\sigma Gq\rangle}{4718592\,\pi^{8}} -\frac{45\,m_s^2\,\langle\bar qq\rangle\,\langle g_s\bar q\sigma Gq\rangle}{16384\,\pi^{6}} +\frac{251\,\langle g_s\bar q\sigma Gq\rangle^{2}}{589824\,\pi^{6}} -\frac{145\,\langle\alpha_sGG\rangle\,\langle\bar qq\rangle\,\langle\bar ss\rangle}{884736\,\pi^{6}} \right.\right. \nonumber\\
&& \left.\left. \qquad -\frac{47\,m_s\,\langle\bar qq\rangle^{2}\,\langle\bar ss\rangle}{9216\,\pi^{4}} -\frac{99\,m_s^2\,\langle g_s\bar q\sigma Gq\rangle\,\langle\bar ss\rangle}{32768\,\pi^{6}} -\frac{11\,\langle\alpha_sGG\rangle\,\langle\bar ss\rangle^{2}}{589824\,\pi^{6}} +\frac{37\,m_s\,\langle\bar qq\rangle\,\langle\bar ss\rangle^{2}}{9216\,\pi^{4}} +\frac{11\,\langle\alpha_sGG\rangle\,m_s\,\langle g_s\bar s\sigma Gs\rangle}{1572864\,\pi^{8}} \right.\right. \nonumber\\
&& \left.\left. \qquad -\frac{91\,m_s^2\,\langle\bar qq\rangle\,\langle g_s\bar s\sigma Gs\rangle}{49152\,\pi^{6}} +\frac{197\,\langle g_s\bar q\sigma Gq\rangle\,\langle g_s\bar s\sigma Gs\rangle}{294912\,\pi^{6}} +\frac{3\,\langle g_s\bar s\sigma Gs\rangle^{2}}{65536\,\pi^{6}} \right) + s\,\left(  +\frac{5\,\langle\alpha_sGG\rangle\,m_s^2\,\langle\bar qq\rangle^{2}}{24576\,\pi^{6}} +\frac{\langle\bar qq\rangle^{4}}{864\,\pi^{2}} \right.\right. \nonumber\\
&& \left.\left. \qquad -\frac{131\,\langle\alpha_sGG\rangle\,\langle\bar qq\rangle\,\langle g_s\bar q\sigma Gq\rangle}{442368\,\pi^{6}} -\frac{m_s\,\langle\bar qq\rangle^{2}\,\langle g_s\bar q\sigma Gq\rangle}{24\,\pi^{4}} -\frac{143\,m_s^2\,\langle g_s\bar q\sigma Gq\rangle^{2}}{73728\,\pi^{6}} +\frac{49\,\langle\alpha_sGG\rangle\,m_s^2\,\langle\bar qq\rangle\,\langle\bar ss\rangle}{73728\,\pi^{6}} \right.\right. \nonumber\\
&& \left.\left. \qquad +\frac{\langle\bar qq\rangle^{3}\,\langle\bar ss\rangle}{16\,\pi^{2}} -\frac{89\,\langle\alpha_sGG\rangle\,\langle g_s\bar q\sigma Gq\rangle\,\langle\bar ss\rangle}{442368\,\pi^{6}} -\frac{25\,m_s\,\langle\bar qq\rangle\,\langle g_s\bar q\sigma Gq\rangle\,\langle\bar ss\rangle}{1728\,\pi^{4}} -\frac{7\,\langle\alpha_sGG\rangle\,m_s^2\,\langle\bar ss\rangle^{2}}{294912\,\pi^{6}} +\frac{29\,\langle\bar qq\rangle^{2}\,\langle\bar ss\rangle^{2}}{864\,\pi^{2}} \right.\right. \nonumber\\
&& \left.\left. \qquad +\frac{7\,m_s\,\langle g_s\bar q\sigma Gq\rangle\,\langle\bar ss\rangle^{2}}{1152\,\pi^{4}} -\frac{89\,\langle\alpha_sGG\rangle\,\langle\bar qq\rangle\,\langle g_s\bar s\sigma Gs\rangle}{442368\,\pi^{6}} -\frac{13\,m_s\,\langle\bar qq\rangle^{2}\,\langle g_s\bar s\sigma Gs\rangle}{864\,\pi^{4}} -\frac{11\,m_s^2\,\langle g_s\bar q\sigma Gq\rangle\,\langle g_s\bar s\sigma Gs\rangle}{4096\,\pi^{6}} \right.\right. \nonumber\\
&& \left.\left. \qquad -\frac{7\,\langle\alpha_sGG\rangle\,\langle\bar ss\rangle\,\langle g_s\bar s\sigma Gs\rangle}{147456\,\pi^{6}} +\frac{13\,m_s\,\langle\bar qq\rangle\,\langle\bar ss\rangle\,\langle g_s\bar s\sigma Gs\rangle}{2304\,\pi^{4}} \right) + \left(  +\frac{89\,\langle\alpha_sGG\rangle\,m_s\,\langle\bar qq\rangle^{3}}{82944\,\pi^{4}} +\frac{23\,m_s^2\,\langle\bar qq\rangle^{4}}{432\,\pi^{2}} \right.\right. \nonumber\\
&& \left.\left. \qquad +\frac{7\,\langle\alpha_sGG\rangle\,m_s^2\,\langle\bar qq\rangle\,\langle g_s\bar q\sigma Gq\rangle}{16384\,\pi^{6}} +\frac{11\,\langle\bar qq\rangle^{3}\,\langle g_s\bar q\sigma Gq\rangle}{1152\,\pi^{2}} -\frac{371\,\langle\alpha_sGG\rangle\,\langle g_s\bar q\sigma Gq\rangle^{2}}{3538944\,\pi^{6}} -\frac{103\,m_s\,\langle\bar qq\rangle\,\langle g_s\bar q\sigma Gq\rangle^{2}}{3072\,\pi^{4}} \right.\right. \nonumber\\
&& \left.\left. \qquad +\frac{23\,\langle\alpha_sGG\rangle\,m_s\,\langle\bar qq\rangle^{2}\,\langle\bar ss\rangle}{27648\,\pi^{4}} -\frac{25\,m_s^2\,\langle\bar qq\rangle^{3}\,\langle\bar ss\rangle}{96\,\pi^{2}} +\frac{71\,\langle\alpha_sGG\rangle\,m_s^2\,\langle g_s\bar q\sigma Gq\rangle\,\langle\bar ss\rangle}{147456\,\pi^{6}} +\frac{89\,\langle\bar qq\rangle^{2}\,\langle g_s\bar q\sigma Gq\rangle\,\langle\bar ss\rangle}{576\,\pi^{2}} \right.\right. \nonumber\\
&& \left.\left. \qquad -\frac{169\,m_s\,\langle g_s\bar q\sigma Gq\rangle^{2}\,\langle\bar ss\rangle}{27648\,\pi^{4}} +\frac{71\,\langle\alpha_sGG\rangle\,m_s\,\langle\bar qq\rangle\,\langle\bar ss\rangle^{2}}{82944\,\pi^{4}} +\frac{91\,m_s^2\,\langle\bar qq\rangle^{2}\,\langle\bar ss\rangle^{2}}{3456\,\pi^{2}} +\frac{643\,\langle\bar qq\rangle\,\langle g_s\bar q\sigma Gq\rangle\,\langle\bar ss\rangle^{2}}{10368\,\pi^{2}} \right.\right. \nonumber\\
&& \left.\left. \qquad +\frac{71\,\langle\alpha_sGG\rangle\,m_s^2\,\langle\bar qq\rangle\,\langle g_s\bar s\sigma Gs\rangle}{221184\,\pi^{6}} +\frac{17\,\langle\bar qq\rangle^{3}\,\langle g_s\bar s\sigma Gs\rangle}{324\,\pi^{2}} -\frac{257\,\langle\alpha_sGG\rangle\,\langle g_s\bar q\sigma Gq\rangle\,\langle g_s\bar s\sigma Gs\rangle}{1769472\,\pi^{6}} \right.\right. \nonumber\\
&& \left.\left. \qquad -\frac{737\,m_s\,\langle\bar qq\rangle\,\langle g_s\bar q\sigma Gq\rangle\,\langle g_s\bar s\sigma Gs\rangle}{27648\,\pi^{4}} +\frac{157\,\langle\bar qq\rangle^{2}\,\langle\bar ss\rangle\,\langle g_s\bar s\sigma Gs\rangle}{2592\,\pi^{2}} +\frac{3\,m_s\,\langle g_s\bar q\sigma Gq\rangle\,\langle\bar ss\rangle\,\langle g_s\bar s\sigma Gs\rangle}{512\,\pi^{4}} \right.\right. \nonumber\\
&& \left.\left. \qquad -\frac{19\,\langle\alpha_sGG\rangle\,\langle g_s\bar s\sigma Gs\rangle^{2}}{1179648\,\pi^{6}} \right) \right)\,e^{-s/M_B^2}\,ds - \left(  -\frac{\langle\alpha_sGG\rangle\,\langle\bar qq\rangle^{4}}{6912\,\pi^{2}} +\frac{19\,\langle\alpha_sGG\rangle\,m_s\,\langle\bar qq\rangle^{2}\,\langle g_s\bar q\sigma Gq\rangle}{18432\,\pi^{4}} \right. \nonumber\\
&& \left. \qquad +\frac{m_s^2\,\langle\bar qq\rangle^{3}\,\langle g_s\bar q\sigma Gq\rangle}{9\,\pi^{2}} -\frac{17\,\langle\alpha_sGG\rangle\,m_s^2\,\langle g_s\bar q\sigma Gq\rangle^{2}}{196608\,\pi^{6}} -\frac{\langle\bar qq\rangle^{2}\,\langle g_s\bar q\sigma Gq\rangle^{2}}{128\,\pi^{2}} -\frac{m_s\,\langle g_s\bar q\sigma Gq\rangle^{3}}{768\,\pi^{4}} \right. \nonumber\\
&& \left. \qquad -\frac{19\,\langle\alpha_sGG\rangle\,\langle\bar qq\rangle^{3}\,\langle\bar ss\rangle}{10368\,\pi^{2}} -\frac{23}{108}\,m_s\,\langle\bar qq\rangle^{4}\,\langle\bar ss\rangle +\frac{7\,\langle\alpha_sGG\rangle\,m_s\,\langle\bar qq\rangle\,\langle g_s\bar q\sigma Gq\rangle\,\langle\bar ss\rangle}{6144\,\pi^{4}} +\frac{11\,m_s^2\,\langle\bar qq\rangle^{2}\,\langle g_s\bar q\sigma Gq\rangle\,\langle\bar ss\rangle}{128\,\pi^{2}} \right. \nonumber\\
&& \left. \qquad -\frac{7\,\langle\bar qq\rangle\,\langle g_s\bar q\sigma Gq\rangle^{2}\,\langle\bar ss\rangle}{384\,\pi^{2}} -\frac{25\,\langle\alpha_sGG\rangle\,\langle\bar qq\rangle^{2}\,\langle\bar ss\rangle^{2}}{20736\,\pi^{2}} +\frac{5}{108}\,m_s\,\langle\bar qq\rangle^{3}\,\langle\bar ss\rangle^{2} +\frac{13\,\langle\alpha_sGG\rangle\,m_s\,\langle g_s\bar q\sigma Gq\rangle\,\langle\bar ss\rangle^{2}}{110592\,\pi^{4}} \right. \nonumber\\
&& \left. \qquad -\frac{71\,m_s^2\,\langle\bar qq\rangle\,\langle g_s\bar q\sigma Gq\rangle\,\langle\bar ss\rangle^{2}}{6912\,\pi^{2}} -\frac{13\,\langle g_s\bar q\sigma Gq\rangle^{2}\,\langle\bar ss\rangle^{2}}{2304\,\pi^{2}} +\frac{25\,\langle\alpha_sGG\rangle\,m_s\,\langle\bar qq\rangle^{2}\,\langle g_s\bar s\sigma Gs\rangle}{55296\,\pi^{4}} +\frac{11\,m_s^2\,\langle\bar qq\rangle^{3}\,\langle g_s\bar s\sigma Gs\rangle}{576\,\pi^{2}} \right. \nonumber\\
&& \left. \qquad -\frac{13\,\langle\alpha_sGG\rangle\,m_s^2\,\langle g_s\bar q\sigma Gq\rangle\,\langle g_s\bar s\sigma Gs\rangle}{294912\,\pi^{6}} -\frac{7\,\langle\bar qq\rangle^{2}\,\langle g_s\bar q\sigma Gq\rangle\,\langle g_s\bar s\sigma Gs\rangle}{384\,\pi^{2}} +\frac{7\,m_s\,\langle g_s\bar q\sigma Gq\rangle^{2}\,\langle g_s\bar s\sigma Gs\rangle}{3072\,\pi^{4}} \right. \nonumber\\
&& \left. \qquad +\frac{13\,\langle\alpha_sGG\rangle\,m_s\,\langle\bar qq\rangle\,\langle\bar ss\rangle\,\langle g_s\bar s\sigma Gs\rangle}{110592\,\pi^{4}} +\frac{13\,\langle\bar qq\rangle\,\langle g_s\bar q\sigma Gq\rangle\,\langle\bar ss\rangle\,\langle g_s\bar s\sigma Gs\rangle}{576\,\pi^{2}} -\frac{13\,\langle\bar qq\rangle^{2}\,\langle g_s\bar s\sigma Gs\rangle^{2}}{2304\,\pi^{2}} \right), \nonumber
\\[5mm]
&& {\cal L}_{\Lambda\Lambda}^{\rm OPE}(s_0,M_B^2) = \hspace*{0pt plus 1filll} \label{app:rhoLL} \\
&&  \int_{4m_s^2}^{s_0} \left(  +\frac{s^7}{11560550400\,\pi^{10}} -\frac{m_s^2\,\,s^6}{103219200\,\pi^{10}} + s^5\,\left(  +\frac{m_s\,\langle\bar ss\rangle}{3686400\,\pi^{8}} \right) + s^4\,\left(  -\frac{\langle\alpha_sGG\rangle\,m_s^2}{7864320\,\pi^{10}} \right) \right. \nonumber\\
&& \left. \qquad + s^3\,\left(  +\frac{\langle\alpha_sGG\rangle\,m_s\,\langle\bar ss\rangle}{589824\,\pi^{8}} +\frac{m_s^2\,\langle\bar ss\rangle^{2}}{18432\,\pi^{6}} \right) + s\,\left(  +\frac{\langle\alpha_sGG\rangle\,m_s^2\,\langle\bar ss\rangle^{2}}{24576\,\pi^{6}} \right) + \left(  +\frac{m_s^2\,\langle\bar qq\rangle^{4}}{18\,\pi^{2}} \right) \right)\,e^{-s/M_B^2}\,ds \nonumber\\
&& - \left(  +\frac{m_s^2\,\langle\bar qq\rangle^{3}\,\langle g_s\bar q\sigma Gq\rangle}{18\,\pi^{2}} -\frac{4}{27}\,m_s\,\langle\bar qq\rangle^{4}\,\langle\bar ss\rangle \right). \nonumber
\end{eqnarray}
\end{widetext}

\bibliographystyle{elsarticle-num}
\bibliography{ref}

\end{document}